\PassOptionsToPackage{paperwidth=210mm,paperheight=297mm,centering,hmargin=3.0cm,vmargin=2.5cm}{geometry}
\documentclass[3p]{elsarticle}
\biboptions{sort&compress}

\usepackage{lineno}
\modulolinenumbers[5]

\journal{arXiv}

\usepackage{textcomp}
\usepackage[]{graphicx}
\graphicspath{{./new_figs}{./Figures/}}
\usepackage{dcolumn}
\usepackage{multirow}
\usepackage{bm}
\usepackage{times}
\usepackage[colorlinks = true,
linkcolor = blue,
urlcolor  = blue,
citecolor = blue,
anchorcolor = blue]{hyperref}
\usepackage{placeins}
\usepackage{amsmath}
\usepackage{amssymb}
\usepackage{threeparttable}

\newcommand{\adisl}{$\langle \bold{a} \rangle$ }
\newcommand{\adislns}{$\langle \bold{a} \rangle \mkern-5mu$ }
\newcommand{\hadisl}{$\langle \bold{a} \rangle/2$ }
\newcommand{\cdisl}{$\langle \bold{c} \rangle$ }

\newcommand{\cdislns}{$\langle \bold{c} \rangle \mkern-5mu$ }

\newcommand{\cadisl}{$\langle \bold{c}+\bold{a} \rangle$ }
\newcommand{\cadislns}{$\langle \bold{c}+\bold{a} \rangle \mkern-5mu$ }
\newcommand{\hcadisl}{$\langle \bold{c}+\bold{a} \rangle/2 $ }

\newcommand{\beginsupplement}{
	\setcounter{section}{0}
	\setcounter{page}{1}
	\setcounter{table}{0}
	\setcounter{figure}{0}
	\setcounter{equation}{0}
	\renewcommand{\thetable}{S\arabic{table}}
	\renewcommand{\theHtable}{Supplementary.\thetable}
	\renewcommand{\thefigure}{S\arabic{figure}}
	\renewcommand{\theHfigure}{Supplementary.\thefigure}
	\renewcommand{\theequation}{S\arabic{equation}}
	\renewcommand{\theHequation}{Supplementary.\theequation}
	\renewcommand{\thesection}{S\arabic{section}}
	\renewcommand{\theHsection}{Supplementary.\thesection}

}

\begin{document}

\begin{frontmatter}

\title{Atomistic modelling of all dislocations and twins in HCP and BCC Ti}
\author[1]{Tongqi Wen}
\author[2]{Anwen Liu}
\author[2]{Rui Wang}
\author[3]{Linfeng Zhang}
\author[2]{Jian Han}
\author[4]{Han Wang\corref{a}}
\cortext[a]{Corresponding author}
\ead{wang\_han@iapcm.ac.cn}
\author[1]{David J. Srolovitz}
\author[2,5]{Zhaoxuan Wu\corref{b}}
\cortext[b]{Corresponding author}
\ead{zhaoxuwu@cityu.edu.hk}

\address[1]{Department of Mechanical Engineering, The University of Hong Kong, Hong Kong SAR, China}
\address[2]{Department of Materials Science and Engineering, City University of Hong Kong, Hong Kong SAR, China}
\address[3]{DP Technology, Beijing, China}
\address[4]{Laboratory of Computational Physics, Institute of Applied Physics and Computational Mathematics, Beijing, China}
\address[5]{Hong Kong Institute for Advanced Study, City University of Hong Kong, Hong Kong SAR, China}

\begin{abstract}
Ti exhibits complex plastic deformation controlled by active dislocation and twinning systems.  Understandings on dislocation cores and twin interfaces are currently not complete or quantitative, despite extensive experimental and simulation studies.  Here, we determine all the core and twin interface properties in both HCP and BCC Ti using a Deep Potential (DP) and DFT. We determine the core structures, critical resolved shear stresses and mobilities of \adislns, \cadislns, \cdisl dislocations in HCP and \(\langle111\rangle/2\) dislocations in BCC Ti.  The \adisl slip consists of slow core migration on pyramidal-I planes and fast migration on prism-planes, and is kinetically limited by cross-slips among them.  This behaviour is consistent with ``locking-unlocking'' phenomena in TEM and is likely an intrinsic property.  Large-scale DFT calculations provide a peek at the screw \cadisl core and glide behaviour, which is further quantified using DP-Ti.  The screw \cadisl is unstable on pyramidal-II planes. The mixed \cadisl is nearly sessile on pyramidal-I planes, consistent with observations of long dislocations in this orientation.  The edge and mixed \cadisl are unstable against a pyramidal-to-basal (PB) transition and become sessile at high temperatures, corroborate the difficulties in \cdislns-axis compression of Ti.  Finally, in BCC Ti, the \(\langle111\rangle/2\) screw has a degenerate core with average glide on \(\{112\}\) planes; the \(\langle111\rangle/2\) edge and mixed dislocations have non-dissociated cores on \(\{110\}\) planes.  This work paints a self-consistent, complete picture on all dislocations in Ti, rationalises previous experimental observations and points to future HRTEM examinations of unusual dislocations such as the mixed and PB transformed \cadisl cores.
\end{abstract}


%

\end{frontmatter}


\section{\label{sec:sec1}Introduction}
Ti has low density, high strength and excellent resistance to corrosion.  It exhibits multiple allotropes which can be tuned through alloying and thermo-mechanical processing.  These combined properties make it attractive for structure material applications.  A wide range of Ti-based alloys have been developed and used extensively in the aerospace~\cite{cotton_2015_jom}, chemical and biomedical industries~\cite{lutjering_2007_ti,banerjee_2013_am}.  For examples, commercially pure HCP-\(\alpha\) Ti is widely used as corrosion resistant materials in the petrochemical industry, while the dual-phase Ti-6Al-4V alloy and its derivatives are among the primary materials in constructing load-carrying aircraft structures and engine components.  BCC-\(\beta\) Ti is also versatile; its alloys can be heat-treated/precipitate-strengthened to very high strength~\cite{boyer_2005_jmep}, or tuned to possess high fatigue strength, low modulus and be bio-compatible~\cite{wang_1996_msea,niinomi_2002_mmta,raabe_2007_am}.

\begin{figure}[!htbp]
	\centering
	\includegraphics[width=0.65\textwidth]{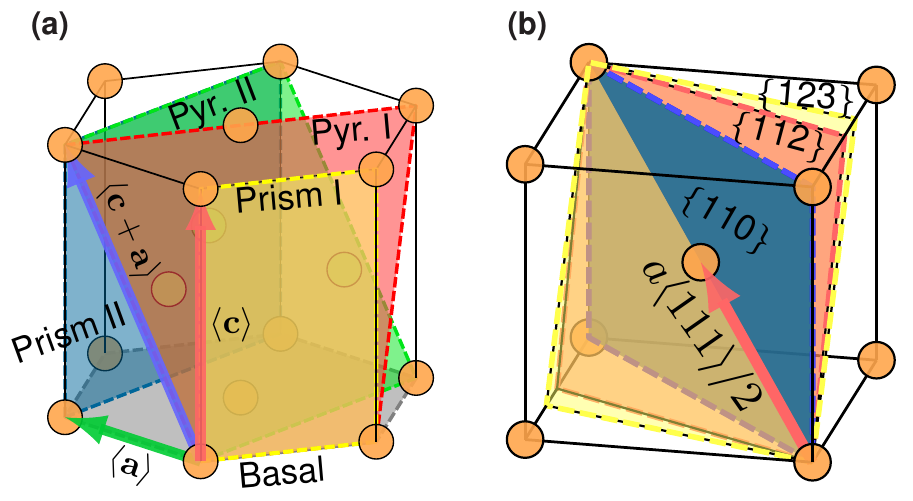}
	\caption{\label{fig:schem_slip} Dislocation slip systems in Ti. (a) HCP structure and (b) BCC structure. }
\end{figure}

The diversity of Ti applications is well-recognized, so too are the complexities associated with its thermodynamic and mechanical properties.  In the HCP structure, Ti exhibits strong elastic~\cite{ogi_2004_am} and plastic anisotropy~\cite{gong_2009_am}.  Its plastic deformation is carried out by dislocation slip and accompanied by deformation twinning at low temperatures.  Slip via dislocation occurs in the close-packed \adisl direction on prismatic, pyramidal I and sometimes basal planes, and in the \cadisl direction on pyramidal I and II planes~\cite{castany_2007_am,yu_2013_sm} (Fig.~\ref{fig:schem_slip}a). There are four primary deformation twinning modes: (i) \(\langle 10\bar{1}1 \rangle \)\(\{10\bar{1}2\}\), (ii) \(\langle  11\bar{2}6 \rangle\)\(\{11\bar{2}1\}\), (iii)\(\langle  11\bar{2} 3 \rangle\)\(\{11\bar{2}2\}\) and (iv) \(\langle  10\bar{1}2\rangle\)\(\{10\bar{1}1\}\) -- the first two modes lead to extension and the latter two to compression in the crystallographic \cdisl direction~\cite{yoo_1981_mta}.  While each deformation mode is intrinsically different, a wide range of experiments show that their activations and operations are sensitive to alloy compositions (e.g. Al~\cite{williams_2002_mmta,zhang_2019_sa} and oxygen content~\cite{zaefferer_2003_msea,yu_2015_science,chong_2020_sa}) and loading conditions (e.g., temperatures and loading direction~\cite{jones_1981_am,gong_2009_am}), making plastic deformation in HCP Ti highly complex.  In addition, HCP Ti exhibits various anomalous behaviours, e.g. shear instability in compression along the \cdisl axis at high temperatures~\cite{williams_2002_mmta} and dynamic strain-ageing within some temperature ranges~\cite{nemat_1999_am}. Such unusual plastic behaviour may be traced to dislocation core properties; these are not well understood and currently under active investigation.

BCC Ti also exhibits complex elastic and plastic behaviour. Below 1155 K, pure BCC Ti is mechanically unstable as its elastic constants violate the Born criterion (\(C_{11} < C_{12}\)~\cite{mouhat_2014_prb}). The BCC structure can be stabilized by entropy~\cite{kadkhodaei_2017_prb} and/or alloying~\cite{li_2007_prl,huang_2016_am} (e.g., V, Mo, Nb, etc).  However, the lack of stability of pure BCC Ti at low temperatures makes it difficult to study the intrinsic dislocation slip behaviour.  In solute-stabilized BCC Ti, plastic deformation is also highly complex; dislocation slip, twinning and martensitic transformation may each be activated, depending on alloy compositions~\cite{kuroda_1998_msea} and loading conditions~\cite{brozek_2016_sm,gao_2018_am}.   In-situ TEM studies suggest that \(\langle  111 \rangle/2\) screw dislocations can glide on \(\{110\}\), \(\{112\}\) and \(\{123\}\) planes (Fig.~\ref{fig:schem_slip}b) and control plastic deformation in some \(\beta\)-alloys, similar to that in BCC transition metals.  However, dislocation behaviour in stable/metastable \(\beta\)-Ti is strongly influenced by the presence of solutes at high concentration (e.g., 23Nb-0.7Ta-2Zr-0.4Si~\cite{castany_2012_sm}).  More fundamentally, dislocation core structures remain mysterious in pure \(\beta\)-Ti, and in particular, the screw core is largely unknown, since it is experimentally intractable~\cite{mendis_2006_pm}.  In addition, some \(\beta\) alloys exhibit unusual plastic behaviour, such as nearly zero work hardening, making them susceptible to plastic instability (e.g., shear localisation during forming).

Given the technological importance and complex plastic behaviour of Ti, extensive theoretical and computational studies have been carried out to clarify its dislocation properties.  In particular, density-functional theory (DFT) calculations have been performed to determine the core structures of the \adisl dislocation on the pyramidal I, prism I and basal planes of HCP Ti~\cite{ghazisaeidi_2012_am,clouet_2015_natmat,poschmann_2017_msmse,kwasniak_2019_sm,tsuru_2022_cms}.  The screw \adisl dislocation exhibits multiple core structures and the ground state is on the pyramidal I plane.  However, the energy differences among the different cores are small, on the order of 10 meV/\AA, and are sensitive to boundary conditions, valence electrons and \(k\)-point mesh used in DFT calculations~\cite{poschmann_2017_msmse}.  Combined with \textit{in-situ} TEM study, the screw \adisl dislocation was shown to cross-slip between the low-energy pyramidal I plane and the high-energy prism I plane, exhibiting a rate-limiting ``locking-unlocking" process during glide~\cite{clouet_2015_natmat}.  This extreme delicacy of the core structure and energy poses significant challenges for empirical/semi-empirical interatomic potentials in accurately reproducing the screw \adisl core structures and glide behaviour~\cite{wen_2021_npjcm,rida_2022_mat}.  Separately, the operation of \cadisl dislocations is critical in Ti to provide the plastic strain accommodation in the crystallographic \cdisl direction (in addition to twinning). However, the \cadisl dislocations have wide core dissociations~\cite{yin_2017_am}, requiring simulation supercells beyond those accessible to routine DFT calculations.  Their core structures and glide behaviour have not been determined with first-principle accuracy.  In the unstable \(\beta\)-Ti,  dislocation core structures cannot be computed at 0 K, so their core properties and glide behaviour remain largely unknown.

While the combined experimental and simulation efforts have shed light on the plastic deformation behaviour in Ti, quantitative understanding is still lacking, particularly on all the relevant dislocation core properties.  In this work, we leverage a recently-developed machine learning (ML) Deep Potential for Ti (DP-Ti~\cite{wen_2021_npjcm}) and large-scale DFT calculations to study all dislocation core structures and mobilities in both the HCP and BCC Ti, painting a complete and self-consistent picture of dislocation properties in Ti.  In particular, we investigate the screw, edge, and/or mixed dislocations (\adislns, \cadislns, \cdisl and \(\langle  111 \rangle/2\)) based on an improved version of DP-Ti~\cite{wen_2021_npjcm} and supplemented by DFT.  All possible core dissociations, relative energies and mobilities are quantitatively determined in HCP and BCC Ti.  In addition, we directly demonstrate the ``locking-unlocking" phenomenon of the screw \adisl dislocation via molecular dynamics (MD) simulations using a large simulation cell and over an extended simulation time.  On the pyramidal I plane, screw \cadisl dislocation core structures are computed and compared in DFT and DP-Ti. Good agreement between these two approaches provides a basis for quantitative measurements of the mobility of \cadisl dislocations. Furthermore, the $\langle 111 \rangle/2$ screw dislocation is shown to possess a degenerate (D) core in BCC Ti at 1000 K, in agreement with prediction based upon a new material index \(\chi\)~\cite{wang_2022_arxiv}.  The screw core is also shown to exhibit complex glide behaviour with a mobility lower than its mixed and edge counterparts.  The current study reveals the intrinsic dislocation properties in Ti and provides a basis for understanding dislocation behaviour in Ti alloys.

\section{\label{sec:method}Simulation Methods and Models}

\subsection{\label{sec:dpti}DP-Ti for Atomistic Modelling}

Atomistic modelling of Ti is challenging as empirical and semi-empirical interatomic potentials have limited capabilities in describing the bonding characteristics in different Ti allotropes (\(\omega\), \(\alpha\) and \(\beta\)).  An MEAM interatomic potential~\cite{hennig_2008_prb} and a recent ML Deep Potential~\cite{wen_2021_npjcm} have been shown to possess the necessary attributes for modelling dislocations in multiphase Ti in a self-consistent manner~\cite{wen_2021_npjcm}.  In particular, these two potentials accurately capture the cohesive energies, elastic constants and generalised stacking fault energies of both the HCP-\(\alpha\) and BCC-\(\beta\) phases.  In the current work, we employ an improved version of the Deep Potential for Ti (DP-Ti).  DP-Ti is trained using the DeePMD-kit~\cite{wang_2018_cpc} through a workflow consisting of ``Initialisation", ``DP-GEN loop", and ``Specialisation" steps.  The detailed training process and benchmarks are described in Ref.~\cite{wen_2021_npjcm}.  Here, we briefly describe the improvements made relative to the earlier version~\cite{wen_2021_npjcm}. The current training datasets include additional \textit{ab initio} MD-based 2$\times$2$\times$2 super cells of the $\omega$-phase.  In the specialisation-step, the training datasets include segments of the $\gamma$-line in the \adisl direction on the \(\{10\bar{1}0\}\)-prism I wide plane and in the \cadisl direction on the pyramidal II plane (Fig.~\ref{fig:schem_slip}).  The fine-tuned DP-Ti shows properties largely similar to that of the previous DP-Ti~\cite{wen_2021_npjcm}.  It also exhibits satisfactory agreement with DFT/experiment results on lattice constants and elastic constants at 0 K (Table.~\ref{tab:ti_properties} in the Supplementary Materials) and finite temperatures (Fig.~\ref{fig:finite_t_lat_els}), equation of states (Fig.~\ref{fig:ti_eos}), \(\gamma\)-lines (Figs.~\ref{fig:gamma_line_hcp} and~\ref{fig:gamma_line_bcc}) and \(\gamma\)-surfaces (Fig.~\ref{fig:gamma_surface_hcp}), as well as phase transition temperatures (Fig.~\ref{fig:ti_phase_diagram}). The details of generating the $\omega$ phase and special datasets, and benchmarks of the current DP-Ti are included in the Supplementary Materials. The current DP-Ti is compatible with the Large-scale Atomic/Molecular Massively Parallel Simulator (LAMMPS~\cite{thompson_2022_cpc}) on both CPU and GPU machines and can thus be readily employed by other interested researchers~\cite{dpti_2022_dplib}. In addition, the DP compression method~\cite{lu_2022_jctc} typically accelerates speeds and reduces memory consumption both by an order of magnitude relative to the original models, enabling large-scale DP-based MD simulations.

\subsection{\label{sec:dft_method}Density Functional Theory Calculations}
In the current work, we employ the DFT method as implemented in the Vienna \textit{ab initio} simulation package (VASP)~\cite{kresse_1996_cms,kresse_1996_prb}. The generalised gradient approximation (GGA) with Perdew-Burke-Ernzerhof (PBE)~\cite{perdew_1996_prl} parameterization is used for the exchange-correlation functional. The 3d\(^2\)4s\(^2\) electrons are treated as valence electrons and the rest as core electrons which are replaced by projector-augmented-wave (PAW~\cite{blochl_1994_prb}) pseudopotentials.  The cut-off energy of the plane-wave basis is 400 eV.  The first order Methfessel-Paxton smearing method~\cite{methfessel_1989_prb} is used for partial electron occupancy (smearing width $\sigma=0.22$ eV).  The Monkhorst-pack \(k\)-point mesh~\cite{monkhorst_1976_prb} is used to sample the Brillouin zone and the linear \(k\)-point spacing is kept at 0.2 $\rm \AA^{-1}$.

To compute the dislocation core of the \cadisl screw dislocation, we first create a fully periodic supercell of ideal HCP lattice with cell vectors \(\mathbf{c}_1 =[1\bar{2}13]/3, \mathbf{c}_2 = 10[10\bar{1}0], \mathbf{c}_3 = 20[1\bar{2}10]/3 \).  The supercell contains 800 atoms (3200 valence electrons). A screw \cadisl dislocation dipole is introduced at \((0.24\mathbf{c}_2, 0.24 \mathbf{c}_3)\) and \((0.76\mathbf{c}_2, 0.76 \mathbf{c}_3)\) by displacing atoms according to the anisotropic linear elastic displacement field of the corresponding Volterra dislocations~\cite{hirth_1992_disl}.  An affine shear deformation is further imposed to compensate the plastic strain induced by the dislocation dipole in the supercell~\cite{bulatov_2006_csd}.  The constructed dislocation cores are optimised using the conjugate gradient algorithm. Convergence is assumed when the energy difference drops below $10^{-4}$ eV between consecutive steps in both the self-consistency electronic and ionic steps.

\subsection{\label{sec:md_core}Molecular Dynamics Simulations}

MD simulations are performed using LAMMPS~\cite{thompson_2022_cpc} through the interface provided in the DeePMD-kit package~\cite{wang_2018_cpc}. We employ cylindrical supercells with a radius \(R\) of 100 {\AA} in the MD dislocation core structure simulations. Dislocations are created at the centre of the cylinder with line directions parallel to the cylinder axis. Table~\ref{tab:md_supercell} summarises the dislocation supercell geometries.  Specifically, we apply the anisotropic linear elastic displacement to all the atoms according to the corresponding Volterra partial dislocations~\cite{hirth_1992_disl}.  Atoms within 2 $\times$ the cutoff distance (2$r_\text{c}$) of the potential from the cylinder outer surface are treated as boundary atoms.  The constructed cores are optimised at 0 K with boundary atoms fixed at the elastic displacement solution.  Since the \cadisl edge dislocation line on the pyramidal I plane does not lie along a rational crystallographic direction, we focus on a mixed \cadisl dislocation aligned in the \adisl direction (frequently observed in experiment~\cite{numakura_1986_sm}).

\begin{table}[!htbp]
	\centering
	\small
  \begin{threeparttable}[b]
	\caption{\label{tab:md_supercell} Crystallographic orientations (\(\mathbf{x},\mathbf{y},\mathbf{z}\)) and approximate simulation cell sizes (\AA) for different dislocations on various planes. For non-pure-edge dislocations, the Burgers vector screw component is always aligned in the \(\mathbf{x}\) direction. The cylindrical supercells have radius \(R\) and length \(t\). The cuboid/parallelepiped supercells have dimensions \(l_1 \times l_2 \times l_3\).  }
  \begin{tabular}{llllll}
    \hline
    Crystal & Burgers vector \& & \multirow{2}{*}{Slip plane} & Crystallographic & Cylinder & Cuboid/Parallelepiped \\
    structure & dislocation type & & orientations & \(t\) \(\times\) \(R\) (\AA) & \(l_1\) \(\times l_2\) \(\times l_3\) (\AA) \\
    \hline
    \multirow{14}{*}{HCP} & \adisl screw & basal & [1\(\bar{2}\)10], [10\(\bar{1}\)0], [0001] & \(2.94 \times 100\) & \(120 \times 300 \times 150\)  \\
    & \adisl screw & prism I & [1\(\bar{2}\)10], [0001], [10\(\bar{1}\)0] & \(2.94 \times 100\) & \(120 \times 300 \times 150\)  \\
    & \adisl screw & pyramidal I & [1\(\bar{2}\)10], [0001], [10\(\bar{1}\)0] & \(2.94 \times 100\) & \(120 \times 300 \times 150\)  \\
    & \adisl edge & basal & [1\(\bar{2}\)10], [10\(\bar{1}\)0], [0001] & \(5.09 \times 100\) & \(300 \times 120 \times 150\) \\
    & \adisl edge & prism I & [1\(\bar{2}\)10], [0001], [10\(\bar{1}\)0] & \(4.64 \times 100\) & \(300 \times 120 \times 150\)\\
    & \adisl edge & pyramidal I & [1\(\bar{2}\)10], [10\(\bar{1}\)2], [\(\bar{1}\)011] & \(10.58 \times 100\) & \(300 \times 120 \times 150\)\\
    \\
   & \cadisl screw & pyramidal I & [1\(\bar{2}\)13], [10\(\bar{1}\)0], [\(\bar{1}\)2\(\bar{1}\)2] & \(5.49 \times 100 \) & \(80 \times 200 \times 100\) \\
   & \cadisl screw & pyramidal II & [1\(\bar{2}\)13], [10\(\bar{1}\)0], [\(\bar{1}\)2\(\bar{1}\)2] & \(5.49 \times 100 \) & \(80 \times 200 \times 100\)\\
   & \cadisl mixed & pyramidal I & [1\(\bar{2}\)10], [10\(\bar{1}\)2], [\(\bar{1}\)011] & \(2.94 \times 100 \) & \(40 \times 430 \times 200\) \\
   & \cadisl edge & pyramidal II & [1\(\bar{2}\)13], [10\(\bar{1}\)0], [\(\bar{1}\)2\(\bar{1}\)2] & \(5.09 \times 100\) & \(200 \times 80 \times 100\)\\
    \\
    & \cdisl screw & prism I & [0001], [1\(\bar{2}\)10], [10\(\bar{1}\)0] & \(4.64 \times 100\) & -\\
    & \cdisl edge & prism I & [0001], [1\(\bar{2}\)10], [10\(\bar{1}\)0] & \(2.94 \times 100\) & -\\
    & \cdisl edge & prism II & [0001], [10\(\bar{1}\)0], [1\(\bar{2}\)10] & \(5.09 \times 100\) & -\\
    & & & & & \\
    \multirow{3}{*}{BCC} & \(a \langle 111 \rangle/2 \) screw & \{110\} & [111], [11\(\bar{2}\)], [1\(\bar{1}\)0] & \(2.85 \times 100\) & \(40 \times 300 \times 150\)\\
            & \(a \langle 111 \rangle/2 \) edge & \{110\} & [111], [11\(\bar{2}\)], [1\(\bar{1}\)0] & \(8.1 \times 100\) & \(300 \times 40 \times 150\)\\
            & \(a \langle 111 \rangle/2 \)  mixed & \{110\} & [111], [11\(\bar{2}\)], [1\(\bar{1}\)0] & \(2.85 \times 100\) & \(40 \times 300 \times 150\) \\
    \hline
  \end{tabular}
  \end{threeparttable}
\end{table}

To compare the relative energies of different core dissociations, we calculate the total excessive energy per unit dislocation length in a cylinder of radius \(r\) from the core centre in the simulation supercell as
\begin{equation}\label{eq:disl_energy_atomistic}
  E_{\rm disl} (r) = E_{\rm total} (r) - n(r) E_{\rm c},
\end{equation}
where \(E_{\rm total} (r)\) and \(n(r)\) are the total potential energy and number of atoms within the cylinder of radius \(r\), and \(E_{\rm c}\) is the cohesive energy per atom in the corresponding perfect crystal.  The total dislocation energy per unit length may also be expressed as
\begin{eqnarray}\label{eq:disl_energy}
	E_{\rm disl} (r) = E_{\rm core}(r_\text{min})+K{\rm ln}(r/r_{\rm min}),
\end{eqnarray}
where $E_{\rm core}(r_\text{min})$ is the near-core energy and $K{\rm ln}(r/r_{\rm min})$ is the elastic energy outside the core region.  The near-core energy includes excessive energies associated with the partial core centre, the stacking fault energy between the partials, their interactions, and thus depends on details of the atomistic structure. \(E_{\rm core}(r_\text{min})\) can only be defined with a specific core radius \(r_{\rm min}\); the choice of \(r_{\rm min}\) is not unique and depends on the core dissociation width. Here, we choose $ r_{\rm min} \in [4b, 6b, 8b]$, where $b$ is the magnitude of the corresponding Burgers vector (Table~\ref{tab:md_supercell}).  At sufficiently large \(r > r_{\rm min}\), the total dislocation energy scales linearly with \(\ln{r}\).  The scaling factor is the energy pre-factor \(K\), which depends on the material elastic constant tensor \(\mathbf{C}\), dislocation Burgers vector \(\mathbf{b}\) and line direction \(\boldsymbol{\xi}\), i.e., \(K(\mathbf{C}, \mathbf{b}, \boldsymbol{\xi})\).  Screw dislocations of the same Burgers vector but different dissociations all have the same \(K\) and the total energy difference thus only depends on the near-core energy (core structure and dissociation).  The relative energies associated with different dissociations/cores can thus be compared without ambiguity.  Edge dislocations can have both different core dissociations and line directions \(\boldsymbol{\xi}\). Their relative energies have clear physical meanings only when \(\boldsymbol{\xi}\) are the same.  In all the cases,  \(K\) can be computed analytically using anisotropic linear elasticity theory~\cite{hirth_1992_disl} or obtained via fitting to \(E_{\rm disl}\) using Eq.~\eqref{eq:disl_energy}.  In the current work, we use both methods for cross-validation.  Finally, the dislocation core structures are visualized using the component of the Nye tensor~\cite{hartley_2005_am} and the differential displacement (DD) map~\cite{vitek_1970_pma}.

\subsection{\label{sec:md_mobility} Critical Resolved Shear Stresses and Dislocation Mobilities}

\begin{figure*}[!htbp]
  \centering
	\includegraphics[width=0.5\textwidth]{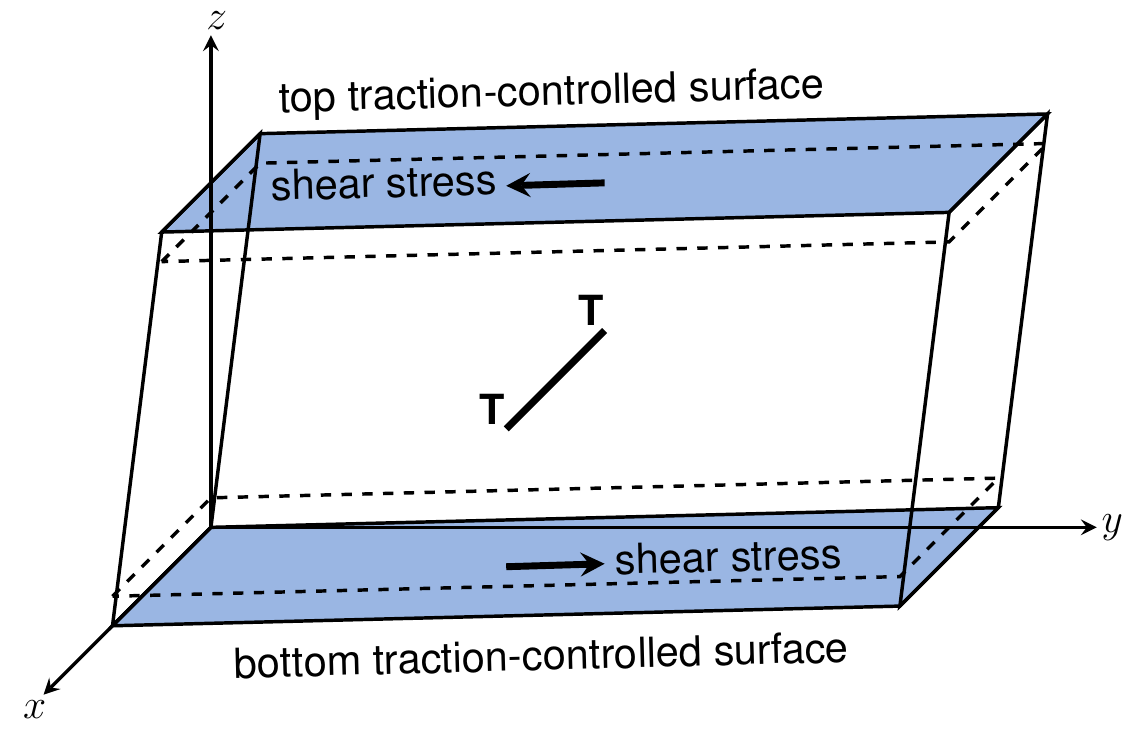}
	\caption{\label{fig:supercell_schem} Supercell for the study of dislocation mobility using the periodic array of dislocations (PADs) configuration. Periodic boundary conditions are imposed in the \(x\) and \(y\) directions. Tractions are applied on atoms on the top and bottom layers within 12 \AA\ from the surfaces in the \(\pm z\) directions. The glide plane normal is \(z\). Supercell vectors are not orthogonal to each other in general. For all non-pure-edge dislocations, the screw component of the Burgers vector is always aligned in the \(x\) direction.}
\end{figure*}

The core structure and energy vary periodically with the crystal periodicity along the glide direction.  The energy variation or landscape consists of low energy valleys and high energy ridges.  The difference between the valley and ridge is the Peierls barrier \(\Delta E_\text{PB}\) which must be overcome during dislocation glide.  At 0 K, the applied stress \(\tau_\text{app}\) provides the driving force and tilts the energy landscape.  With increasing \(\tau_\text{app}\),  dislocations may start to glide when \(\Delta E_\text{PB}\) is zero and \(\tau_\text{app}\) can be taken as the critical resolved shear stress (CRSS) \(\tau_\text{PS}\) at 0 K.  In the current work, if \(\tau_\text{PS}\) is finite, the dislocation is viewed as glissile.  If other plastic deformation modes, such as nucleation of other dislocations, cross-slip to other planes, or fracture, occur before dislocation glide, the dislocation can be viewed as sessile in the current configuration.  The screw \adisl on the pyramidal I plane is special in the sense that it cross-slips to the easy-glide, prism I plane and is considered to be sessile on the pyramidal I plane at 0 K under \(\tau_\text{app}\).

  At finite temperatures, the Peierls barrier is overcome by a combination of thermal activation and applied stress.  The activation energy under stress can be expressed as
  \begin{equation}
    \Delta E_\text{B} = b \int (\tau_\text{PS} - \tau_\text{app}) \text{d}A,
  \end{equation}
  where $b$ is the magnitude of the Burgers vector and the integration is carried over the activation area $A$.  When \(\tau_\text{app} < \tau_\text{PS}\), the dislocation velocity is generally written as
  \begin{equation} \label{eq:disl_kink_nucleation}
    v = \nu_0\exp{\left( - \dfrac{\Delta E_\text{B} }{kT}  \right)},
  \end{equation}
  where \(\nu_0\) is the attempt frequency.  Dislocation glide is thus stochastic and depends on \(\tau_\text{app}\) or wait time in the measurement. At finite \(\tau_\text{app} < \tau_\text{PS}\) and infinite wait time, dislocation glide is dominated by the thermal activation to nucleate a critical kink and kink propagation.  A dislocation will eventually attain a finite velocity at all finite \(\tau_\text{app}\).  The CRSS is thus not well-defined; the 0 K definition of \(\tau_\text{PS} \) approaches zero at finite temperatures.  At \(\tau_\text{app} > \tau_\text{PS}\), dislocation glide is limited by phonon drag.  In this regime, we measure dislocation mobilities by recording core positions as a function of time under a constant \(\tau_\text{app}\). The dislocation velocity \(v\) is the time derivative of the core position. The dislocation mobility $m$ is fitted~\cite{olmsted_2005_msmse} as
  \begin{equation}\label{eq:mobility}
    v = b(\tau_\text{app}-\tau_0)m, \,\,\,\, \tau_\text{app} \geq \tau_0,
  \end{equation}
  where $b$ is the magnitude of the Burgers vector, $\tau_\text{app}$ is the applied shear stress in the direction of the Burgers vector on the slip plane, and $\tau_0$ is viewed as the CRSS of the dislocation above which the glide is athermal.  We emphasise that \(\tau_0\) is different from the \(\tau_\text{PS}\) at zero K. For many dislocations in pure materials, \(\tau_0\) is relatively small and rarely controls plasticity.  In the current work, we focus on the regime where \(\tau_\text{app} \geq \tau_0\), \(\tau_0\) and the mobility can be well-established using Eq.~\ref{eq:mobility}.

In practice, dislocation mobilities are measured using a supercell consisting of a periodic array of dislocations (PADs~\cite{bulatov_2006_csd}).  Specifically, we create parallelepiped supercells (Table~\ref{tab:md_supercell}) and impose periodic boundary conditions in the \(x\) and \(y\) directions (Fig.~\ref{fig:supercell_schem}). In the \(z\) direction, traction-controlled surface boundary conditions are applied.  Dislocations are introduced by applying the displacement field of the corresponding Volterra dislocations at the centre of the supercell.  For non-pure-edge dislocations, the screw component \(\mathbf{b}_\text{s}\) of the Burgers vector is always aligned in the \(x\) direction and a homogeneous strain \(\varepsilon_{yx} = \lvert \mathbf{b}_\text{s}\rvert/2\) is imposed on the supercell to account for the plastic strain associated with the screw component.  All constructed supercells with dislocations are first optimised at 0 K and then equilibrated at target temperatures within an isothermal-isobaric NPT ensemble~\cite{shinoda_2004_prb}.  The normal stresses are maintained at zero in the $x$ and $y$ directions during the equilibration.  To drive the dislocation, a shear stress is created by adding forces in opposite directions along the dislocation Burgers vector on surface layer atoms within 12 \AA\ from the top and bottom surfaces.  For the \adisl and \cadisl dislocations in HCP Ti and the \(\langle 111 \rangle/2\) dislocations in BCC Ti, the applied stresses are first ramped with a step of 5 MPa and 10 MPa and equilibrated for 20 ps between each step increment.  The stress-ramping and equilibration process creates various configurations under near-constant stress conditions.   For the screw \adisl dislocation on the pyramidal I plane, a constant stress of 20 MPa is applied to study the ``locking-unlocking" process.

\section{\label{sec:hcp} Dislocations in HCP-\(\alpha\) Ti}

We first present the core properties of the \adislns, \cadisl and \cdisl dislocations in HCP-\(\alpha\) Ti and compare the results with DFT calculations and relevant experimental data.

\subsection{\label{sec:hcp_core}Dislocation Core Structures and Energies}

\begin{figure*}[!htbp]
	\includegraphics[width=1.0\textwidth]{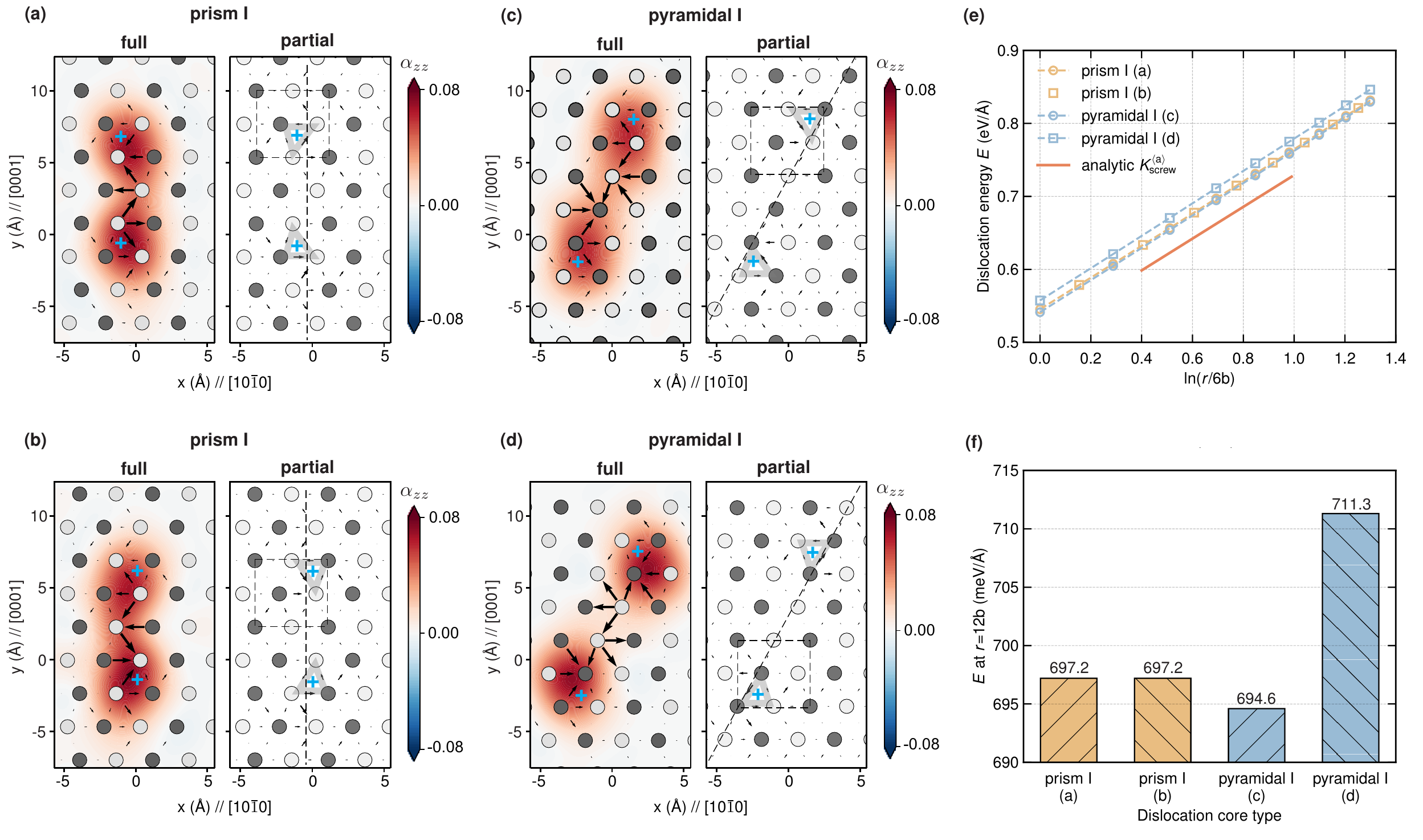}
	\caption{\label{fig:screw_a_hcp}  Screw \adisl dislocation core structures and energies in HCP Ti.  (a-b) Two configurations on the prism I planes. (c-d) Two configurations on the pyramidal I planes. The cores are visualized by the Nye tensor component (\(\alpha_{zz}\), \AA\(^{-1}\)) and differential displacement vectors. The ``\(+\)'' indicates the partial core position revealed in the DD map calculated by using \(\left\vert \mathbf{b} \right\vert =a/2\).  (e) Dislocation energies (a-d) as a function of \(\ln{(r/6b)}\). The slope \(K_{\rm screw}^{\langle a \rangle}\) is calculated using anisotropic linear elasticity theory~\cite{hirth_1992_disl}. (f) Dislocation energies (a-d) measured at \(r = 12b\). Atom colours (dark and light gray) indicate the AB layers along the viewing \adislns direction.}
\end{figure*}

Figure~\ref{fig:screw_a_hcp} shows the core structures and energies of the screw \adisl dislocation dissociated on prism I wide and pyramidal I narrow planes.  The DD maps are plotted in two variants: \(\left\vert \mathbf{b} \right\vert = a \) and \(\left\vert \mathbf{b} \right\vert = a/2 \), to reveal the partial core positions and splitting.  The dissociated screw \adisl cores are similar to the corresponding cores in DFT~\cite{clouet_2015_natmat}.   For all these cases, the energy pre-factors \(K_{\rm screw}\) are identical (within numerical accuracy) and equal to 0.218 eV/\AA\ based on the anisotropic linear elastic theory (Table~\ref{tab:elastic_k}).  On the prism I plane (Figs.~\ref{fig:screw_a_hcp}a and b), the two cores have nearly identical energies (\(< 0.06 \text{ meV}/\)\AA) as measured from the atomistic calculations (Eq.~\eqref{eq:disl_energy_atomistic} and Figs.~\ref{fig:screw_a_hcp}e and f), while the two cores have relatively large energy differences (16.7 meV/\AA) on the pyramidal I plane. The screw \adisl partials on the basal \{0001\} plane appear to be unstable, spontaneously switching to a pyramidal I plane during structure optimisation at 0 K.  This agrees well with direct DFT calculations~\cite{kwasniak_2019_sm}.

\begin{table}[!htbp]
	\centering
	\small
  \begin{threeparttable}[b]
    \caption{\label{tab:elastic_k} The energy pre-factor \(K\) in Eq.~\eqref{eq:disl_energy} computed analytically using anisotropic linear elasticity theory~\cite{hirth_1992_disl} and elastic properties of DP-Ti in comparison with values from linear fitting to \(E_\text{disl}(r)\) measured in atomistic simulations.}
    \begin{tabular}{p{5cm}p{3cm}p{2.5cm}p{2.5cm}}
      \hline
      Burgers vector \& dislocation type & Slip plane & \(K_\text{analytic}\) (eV/{\AA}) & \(K_\text{fitting}\) (eV/{\AA}) \\
      \hline
      \adisl screw & prism I & 0.218 & 0.221\\
      \adisl screw & pyramidal I & 0.218 & 0.222\\
      \adisl edge & basal & 0.320 & 0.326\\
      \adisl edge & prism I & 0.296 & 0.305\\
      \adisl edge & pyramidal I & 0.302 & 0.309\\
      \\
      \cadisl screw & pyramidal I & 0.779 & 0.810\\
      \cadisl screw & pyramidal II & 0.779 & 0.800\\
      \cadisl mixed & pyramidal I & 1.126 & 1.105\\
      \cadisl edge & pyramidal II & 1.152 & 1.155\\
      \\
      \cdisl screw & prism I & 0.587 & 0.599\\
      \cdisl edge & prism I & 0.830 & 0.794\\
      \cdisl edge & prism II & 0.830 & 0.793\\
      \hline
    \end{tabular}
  \end{threeparttable}
\end{table}

\begin{figure*}[!htbp]
  \centering
 	\includegraphics[width=1.0\textwidth]{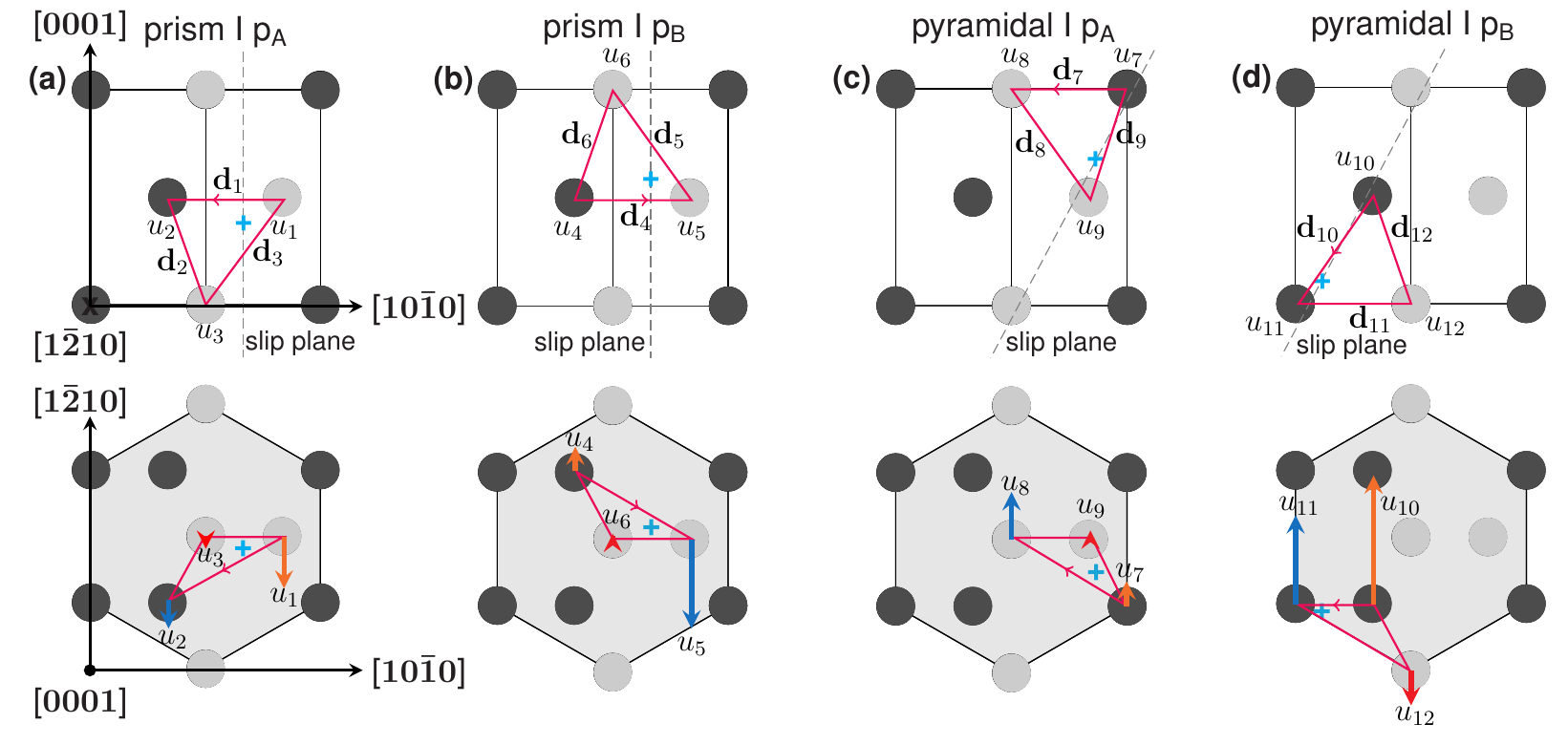}
  \caption{\label{fig:screw_a_disp_hcp} Atom displacements enclosing the various partial screw \adisl cores in Fig.~\ref{fig:screw_a_hcp}. In the bottom row figures, the arrows show the atomic displacements in the Burgers vector \adisl direction measured in molecular static simulations using DP-Ti. (a-b) Partial dislocation cores on the prism I plane. (c-d) Partial dislocation cores on the pyramidal I plane.  Atom colours (dark and light gray) indicate the AB layers along the viewing \adisl direction. }
\end{figure*}

We analyse the screw \adisl partial cores in the HCP structure using fractional Burgers vectors similar to that in BCC screw dislocation cores~\cite{hirth_1992_disl}. At each core centre, the partial Burgers vector is resolved through fractional Burgers vectors as
\begin{align}
  \mathbf{b} = \sum_{ij}  \left( \mathbf{u}_{ij} - \mathbf{U}_{ij} \right)=\sum_{k=1}^3 \mathbf{d}_k = \mathbf{d}_1 + \mathbf{d}_2 + \mathbf{d}_3,
\end{align}
where \(\mathbf{u}_{ij}\) and \(\mathbf{U}_{ij}\) are the displacement vectors between neighbouring atoms \(i\) and \(j\) before and after the introduction of the dislocation and \(\mathbf{d}_k\) is the DD between atoms \(i\) and \(j\). The summation is carried out along a path enclosing the dislocation core.  Figure~\ref{fig:screw_a_disp_hcp} shows the atom displacements at the core centres of the respective partial cores in Fig.~\ref{fig:screw_a_hcp}.  In the first configuration on the prism I plane (Figs.~\ref{fig:screw_a_hcp}a and~\ref{fig:screw_a_disp_hcp}a-b), the two partial cores (prism I \(\text{p}_\text{A}\) and prism I \(\text{p}_\text{B}\)) are different; the partial Burgers vectors are resolved as
\begin{align}
  \mathbf{b}^{\text{prism I p}_\text{A}} &= \mathbf{d}_1 + \mathbf{d}_2 + \mathbf{d}_3; \\
  \mathbf{b}^{\text{prism I p}_\text{B}} &= \mathbf{d}_4 + \mathbf{d}_6 + \mathbf{d}_5,
\end{align}
where the individual measured \(\mathbf{d}_i\) projected in the Burgers vector direction are shown in Table~\ref{tab:dd_magnitude}. In particular, the differential displacement pairs (\(\mathbf{d}_1\) and \(\mathbf{d}_4\), \(\mathbf{d}_2\) and \(\mathbf{d}_6\), \(\mathbf{d}_3\) and \(\mathbf{d}_5\)) measure the DD between atoms at similar positions in the partial cores.  The two partial cores thus exhibit subtle differences in their structures, with each corresponding projected \(\mathbf{d}_i\) differing slightly in magnitudes.  The second configuration on the prism I plane (Fig.~\ref{fig:screw_a_hcp}b) has nearly identical DDs and thus total energy as the first configuration (Figs.~\ref{fig:screw_a_disp_hcp}a-b). We note that it is also possible to obtain configurations with nearly identical partial cores, i.e., \(\mathbf{b}^{\text{prism I p}_\text{A}} \approx \mathbf{b}^{\text{prism I p}_\text{B}}\) in the dislocation dipole configuration.  Therefore, the partial dissociations exhibit multiple states which are sensitive to the supercell boundary conditions on the prism I plane.  Nevertheless, the energy differences are very small among all the core dissociations on the prism I plane.

\begin{figure*}[!htbp]
	\includegraphics[width=1.0\textwidth]{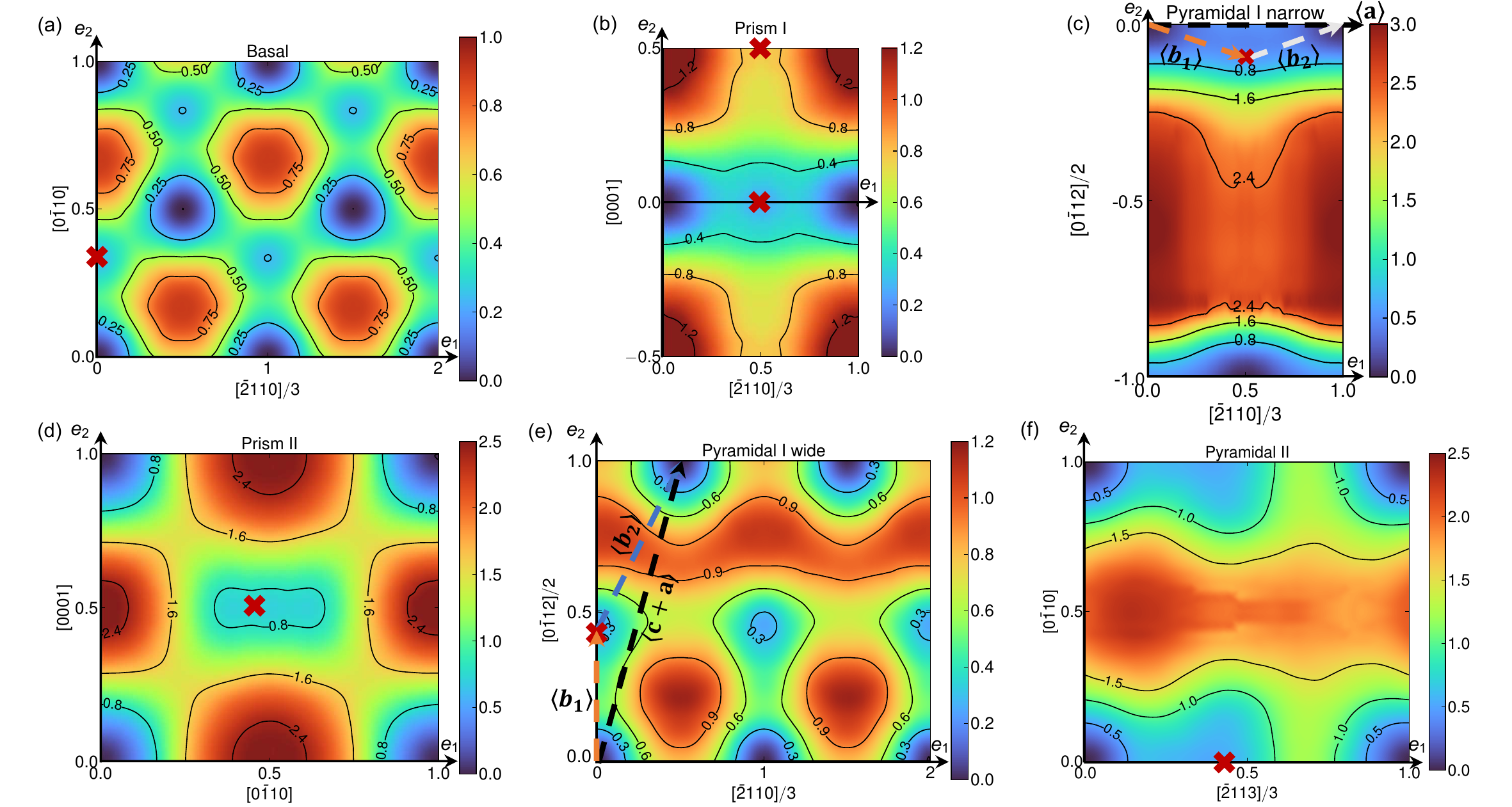}
	\caption{\label{fig:gamma_surface_hcp} The generalised stacking fault energy ($\gamma$) surface of HCP Ti calculated by DP-Ti. The red crosses show all the metastable stacking fault positions from DFT in Table~\ref{tab:ti_hcp_sfs}. The expected dissociations of the \adisl and \cadisl dislocations are indicated by the dashed arrows in (c) and (e), respectively.}

\end{figure*}

On the pyramidal I plane, the screw \adisl dislocation has at least two core dissociations with the lowest and highest energies (Figs.~\ref{fig:screw_a_hcp}c and d), respectively.  Both cores dissociate into two partials with some edge components, as indicated in the \(\gamma\)-surface of the pyramidal I plane (Fig.~\ref{fig:gamma_surface_hcp}).  In both configurations (Figs.~\ref{fig:screw_a_hcp}c and d), the two partial cores are identical and the partial Burgers vectors are resolved as
\begin{align}
  \mathbf{b}^{\text{pyramidal I p}_\text{A}} &= \mathbf{d}_7 + \mathbf{d}_8 + \mathbf{d}_9,
\end{align}
and
\begin{align}
  \mathbf{b}^{\text{pyramidal I p}_\text{B}} &= \mathbf{d}_{11} + \mathbf{d}_{12} + \mathbf{d}_{10},
\end{align}
where the measured fractional Burgers vectors projected in \(\mathbf{b}\) are shown in Table~\ref{tab:dd_magnitude}.

On the pyramidal I plane, the two partial dissociations are thus distinctly different.  Each dissociation has its preferred partial core.  Their total energy difference between the two dissociations is thus expected to be larger than that of the two cores on the prism I plane which have similar partial cores.  Since glide of the \adisl dislocation passes through these two disparate configurations on the pyramidal I plane, the lattice friction is thus expected to be relatively large as well.  This is consistent with direct measurement of the near-core energy (Fig.~\ref{fig:screw_a_hcp}f) and the Peierls stresses of the respective cores (see below).

Comparing the fractional Burgers vectors for the dissociations on the prism and pyramidal planes, we observe that \(\mathbf{d}_8\) of p\(_\text{A}\) on the pyramidal I plane is very similar to that of the \(\mathbf{d}_3\) of p\(_\text{A}\) on the prism I plane, while \(\mathbf{d}_{9}\) of p\(_\text{A}\) on the pyramidal I plane is very similar to \(\mathbf{d}_2\) of p\(_\text{A}\) on the prism I plane.  The DDs of the low energy partial cores on the pyramidal I plane (Fig.~\ref{fig:screw_a_hcp}c) are thus similar to the p\(_\text{A}\) core on the prism I plane (Fig.~\ref{fig:screw_a_hcp}a).  However, the DDs of the p\(_\text{B}\) core on the pyramidal I plane are quite different from the partial cores on the prism I plane. This further suggests that the screw \adisl dislocation dissociation may be controlled by the respective partial core structures and energies, in addition to the metastable stacking fault energies of the prism I and pyramidal I planes.

\begin{table}[!htbp]
	\centering
  \begin{threeparttable}[b]
    \small
    \caption{\label{tab:dd_magnitude} The magnitudes of the differential displacements (DDs) or fractional Burgers vectors in the \adisl direction for the screw \adisl dislocation partial cores. See Fig.~\ref{fig:screw_a_disp_hcp} for the definition of each DD. }
  \begin{tabular}{p{4.8cm}p{2.1cm}p{2.1cm}p{2.1cm}p{2.1cm}}
    \hline
  & prism I p\(_\text{A}\) &  prism I p\(_\text{B}\) & pyramidal I p\(_\text{A}\) &  pyramidal I p\(_\text{B}\)\\
    \hline
    \multirow{3}{*}{ \(\mathbf{d}_i\) projected along \(\mathbf{m} = \mathbf{b}/\lvert \mathbf{b}\rvert\)} & \(\mathbf{d}_1 \cdot \mathbf{m}\): -0.158$a$& \(\mathbf{d}_4 \cdot \mathbf{m}\): -0.147$a$& \(\mathbf{d}_7 \cdot \mathbf{m}\): -0.155$a$& \(\mathbf{d}_{11} \cdot \mathbf{m}\):  -0.163$a$ \\
  &  \(\mathbf{d}_2 \cdot \mathbf{m}\): -0.122$a$& \(\mathbf{d}_6 \cdot \mathbf{m}\): -0.114$a$& \(\mathbf{d}_8 \cdot \mathbf{m}\): -0.221$a$& \(\mathbf{d}_{12} \cdot \mathbf{m}\): -0.106$a$ \\
  &  \(\mathbf{d}_3 \cdot \mathbf{m}\): -0.220$a$& \(\mathbf{d}_5 \cdot \mathbf{m}\): -0.239$a$ & \(\mathbf{d}_9  \cdot \mathbf{m}\): -0.124$a$& \(\mathbf{d}_{10} \cdot \mathbf{m}\): -0.231$a$\\
      Total Burgers vector \(\lvert \mathbf{b} \rvert = \lvert \sum \mathbf{d}_i \cdot \mathbf{m} \rvert \) & -$a$/2 & -$a$/2 & -$a$/2 & -$a$/2\\
      \hline
    \end{tabular}
  \end{threeparttable}
\end{table}
Table~\ref{tab:delta_E_screw_a_hcp} shows the energy differences for all cases and comparison with DFT calculations~\cite{clouet_2015_natmat,poschmann_2017_msmse,tsuru_2022_cms}. The current DP-Ti exhibits the same core energy ordering between the prism and pyramidal dissociations as that in DFT.  Quantitatively, DP-Ti under-estimates the energy differences between the low energy pyramidal I dissociation and the low energy prism I dissociation, as well as the energy difference between the two dissociations on the prism I plane. It also over-estimates the energy difference between the high energy dissociation on pyramidal I plane and the high energy dissociation on the prism I plane.  Nevertheless, these energy differences are quite small; the corresponding values also vary amongst DFT calculations~\cite{clouet_2015_natmat,poschmann_2017_msmse,tsuru_2022_cms}.

\begin{table}[!htbp]
	\centering
  \begin{threeparttable}[b]
	\small
	\caption{\label{tab:delta_E_screw_a_hcp} Energy differences of the screw \adisl dislocation dissociations in Fig.~\ref{fig:screw_a_hcp}.}
  \begin{tabular}{p{5cm}p{2cm}p{2cm}p{2cm}p{2cm}}
    \hline
    Energy difference (meV/\AA) & DP-Ti & DFT~\cite{clouet_2015_natmat} & DFT~\cite{poschmann_2017_msmse} & DFT~\cite{tsuru_2022_cms} \\
			\hline
      \(E_{\rm prism\ I}^{\rm low} - E_{\rm pyramidal\ I}^{\rm low} \) & 2.6 & 5.7 & 6.3 & 7.8\\
      \(E_{\rm prism\ I}^{\rm high} - E_{\rm prism\ I}^{\rm low} \) & \(<0.06\) & \(\sim0.4\) & - & 2.7\\
      \(E_{\rm pyramidal\ I}^{\rm high} - E_{\rm prism\ I}^{\rm high}\) & 14.1 & \(\sim5.3\) &- & 7.2 \\
      \hline
		\end{tabular}
  \end{threeparttable}
\end{table}

On the pyramidal I plane, the lowest and highest energy core structures in DP-Ti are opposite to that in DFT~\cite{clouet_2015_natmat}.  This discrepancy arises from the deficiency of DP-Ti to accurately capture the energy differences between the individual partial cores (similar to the hard and easy cores in BCC transition metals~\cite{clouet_2021_crp}).  This incorrect ordering should not affect pyramidal I plane glide kinetics, since the total energy difference between them are similar to that in DFT (16.7 meV/\AA\ vs. 11-18 meV/\AA). On the prism I plane, the two partial dissociations have very close energies and the kinetic barrier to prism I glide is expected to be relatively low (see below).  Between the pyramidal I and prism I planes, cross-slip is necessary and the kinetic rate for cross-slips may be studied using the current DP-Ti potential.  We demonstrate this in the study of ``locking-unlocking" phenomena which involves the core transition between the prism I and pyramidal I planes.

\begin{figure*}[!htbp]
    \includegraphics[width=1.0\textwidth]{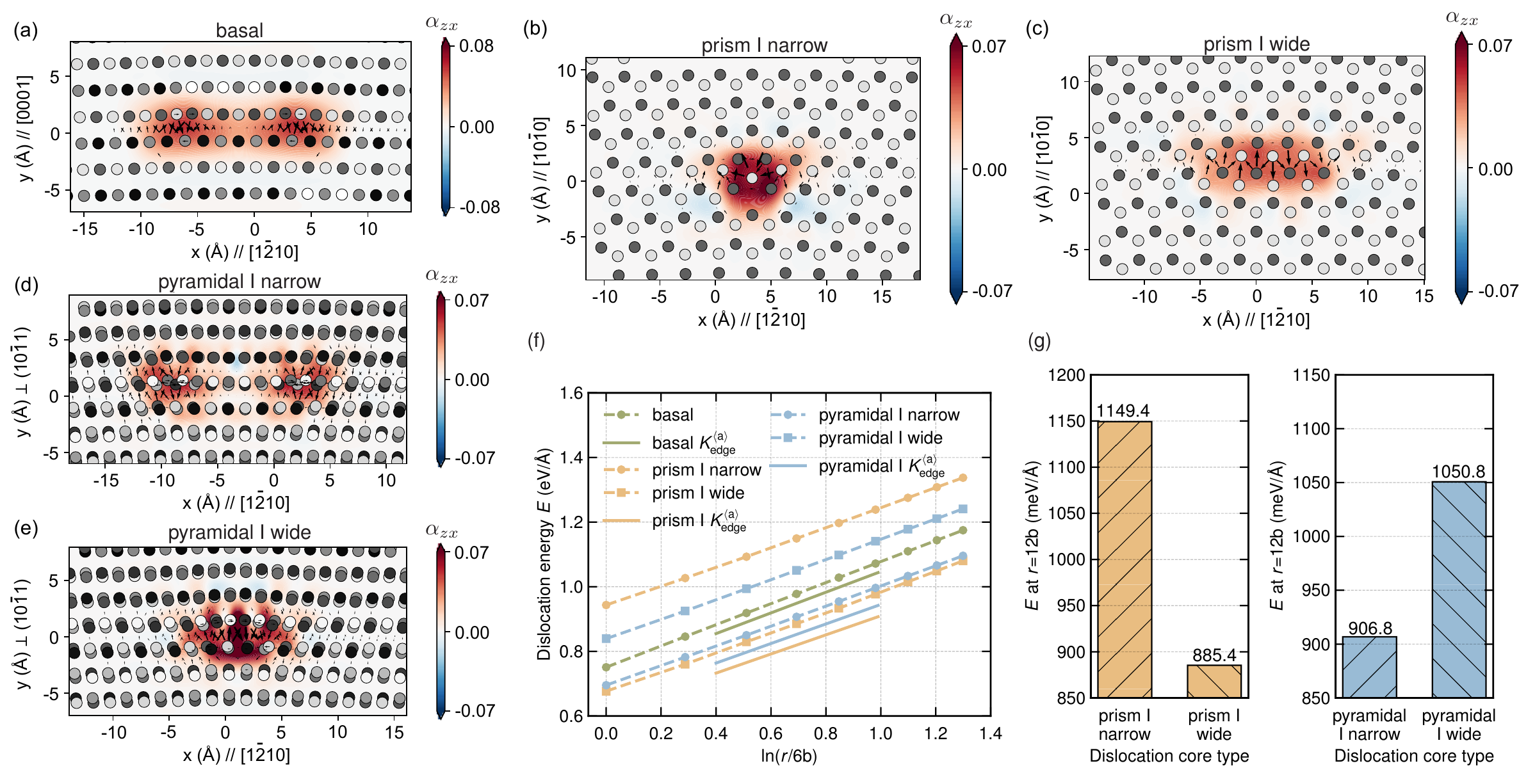}
    \caption{\label{fig:edge_a_hcp} Edge \adisl dislocation core structures and energies in HCP Ti.  (a-e) Edge \adisl dislocation core structures visualized using the Nye tensor (\(\alpha_{zx}\), \AA\(^{-1}\)) and differential displacement plots. (f) Edge \adisl dislocation energy as a function of \(\ln{(r/6b)}\) from Eq.~\eqref{eq:disl_energy}. (g) Energies of edge \adisl dislocations on prism I narrow and wide planes and pyramidal I narrow and wide planes at \(r = 12b\).}
\end{figure*}

Figure~\ref{fig:edge_a_hcp} shows the dislocation core structures and energies of the edge \adisl dislocation dissociated on the basal, prism I (wide and narrow), and pyramidal I (wide and narrow) planes. On the basal plane (Fig.~\ref{fig:edge_a_hcp}a), the \adisl dislocation dissociates into two partials, similar to the Shockley partial pair in FCC structures.  The edge \adisl dislocation should be able to glide on the basal plane, even if its screw counterpart is not stable on this plane.  Planar dissociations are observed on the prism I wide and pyramidal I narrow planes (Figs.~\ref{fig:edge_a_hcp}c, d), while compact cores are seen on the prism I narrow and pyramidal I wide planes (Figs.~\ref{fig:edge_a_hcp}b, e), all consistent with the existence of metastable stacking faults in the \adisl direction on these planes (Fig.~\ref{fig:gamma_surface_hcp}).  In Fig.~\ref{fig:edge_a_hcp}f, the energy pre-factors (\(K_{\rm edge}\)) measured from atomistic calculations are in good agreement with their respective analytic values (Table~\ref{tab:elastic_k}).  We note that the energy pre-factors of the edge dislocations \(K_{\rm edge}\) are different because their line directions \(\boldsymbol{\xi}\) differ from each other with respect to the HCP unit cell.  Therefore, we only compare the energy differences between cores of the same line direction.  In this case, the dissociated cores always have lower energies on the prism I wide and pyramidal I narrow planes, i.e., \(E^{\rm narrow}_{\rm prism\,I} - E^{\rm wide}_{\rm prism\,I} =\) 264 meV/\AA\ and \(E^{\rm wide}_{\rm pyramidal\,I} - E^{\rm narrow}_{\rm pyramidal\,I} =\) 144 meV/\AA\ (Fig.~\ref{fig:edge_a_hcp}g).

\begin{figure}[!htbp]
	\centering
	\includegraphics[width=0.9\textwidth]{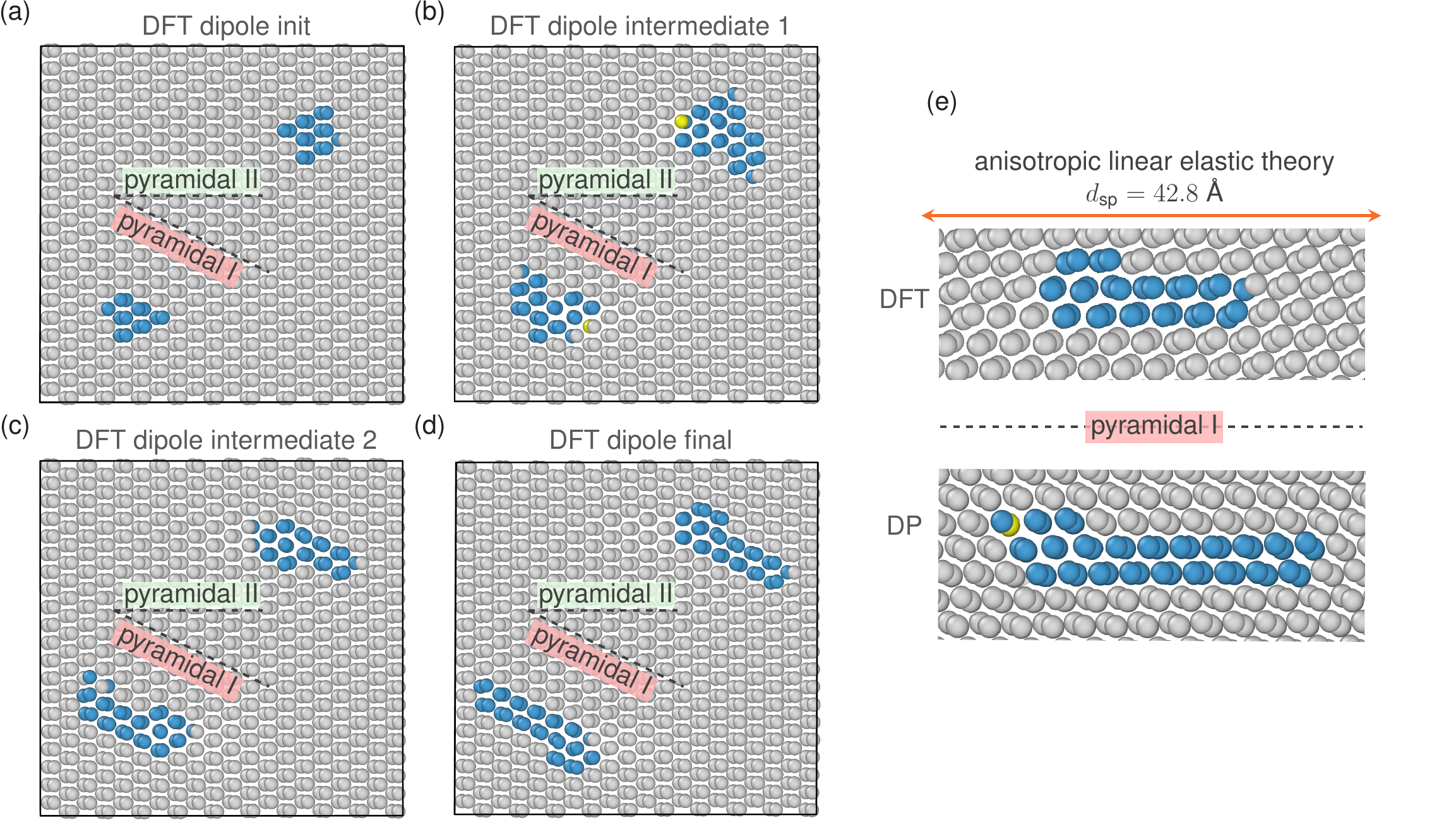}
	\caption{\label{fig:ca_screw_hcp} Screw \cadisl dislocation core structures in HCP Ti.  (a-d) Screw \cadisl dipole dislocation configurations during structure optimisation in DFT. (e) Comparison of the screw \cadisl dislocation core structure calculated by DFT and DP-Ti.  Atoms are coloured based on their local atomic environment/structure as identified by the common neighbour analysis (CNA~\cite{faken_1994_cms}); HCP: grey, BCC: yellow, others: blue. The partial dislocation separation distance \(d_\text{sp}\) is calculated by anisotropic linear elasticity theory~\cite{yin_2017_am}.}
\end{figure}

Figures~\ref{fig:ca_screw_hcp}a-d show the \cadisl dislocation core evolution in the dipole configuration computed by DFT.  During structure optimisation, the initial compact cores spontaneously dissociate onto the pyramidal I planes, forming two pairs of partials with stacking faults on the pyramidal I planes. The two partials have different core structures, as expected from crystal symmetry.  The dissociation process occurs via the glide of the partials in opposite directions along the pyramidal I slip plane as driven by the elastic repulsion between the partial pairs.  The dissociation continues with increasing stacking fault widths, driving the two partial from the two \cadisl towards each other at the periodic boundary.  While the DFT calculations do not converge to a stable dipole configuration due to the current limited supercell (3200 valence electrons), it reveals several critical screw \cadisl core properties.  First, the screw \cadisl is highly favoured on the pyramidal I planes with a well-defined stacking fault, similar to that of HCP Mg~\cite{itakura_2016_prl}. Second, both partial cores have relatively compact core centres. Third, both partials are glissile on pyramidal I planes.  Figure~\ref{fig:ca_screw_hcp}e shows comparisons of the \cadisl screw cores calculated by DFT and DP-Ti. The DP-Ti core is very similar to that in DFT; its partial separation appears to be wider but the DFT core is not in an equilibrium configuration. The DP-Ti partial core separation distance is also shorter than the result of anisotropic linear elasticity theory~\cite{yin_2017_am}.   This discrepancy is largely due to the higher stacking fault energy of DP-Ti on the pyramidal I plane (see Supplementary Materials).

\begin{figure*}[!htbp]
	\includegraphics[width=1.0\textwidth]{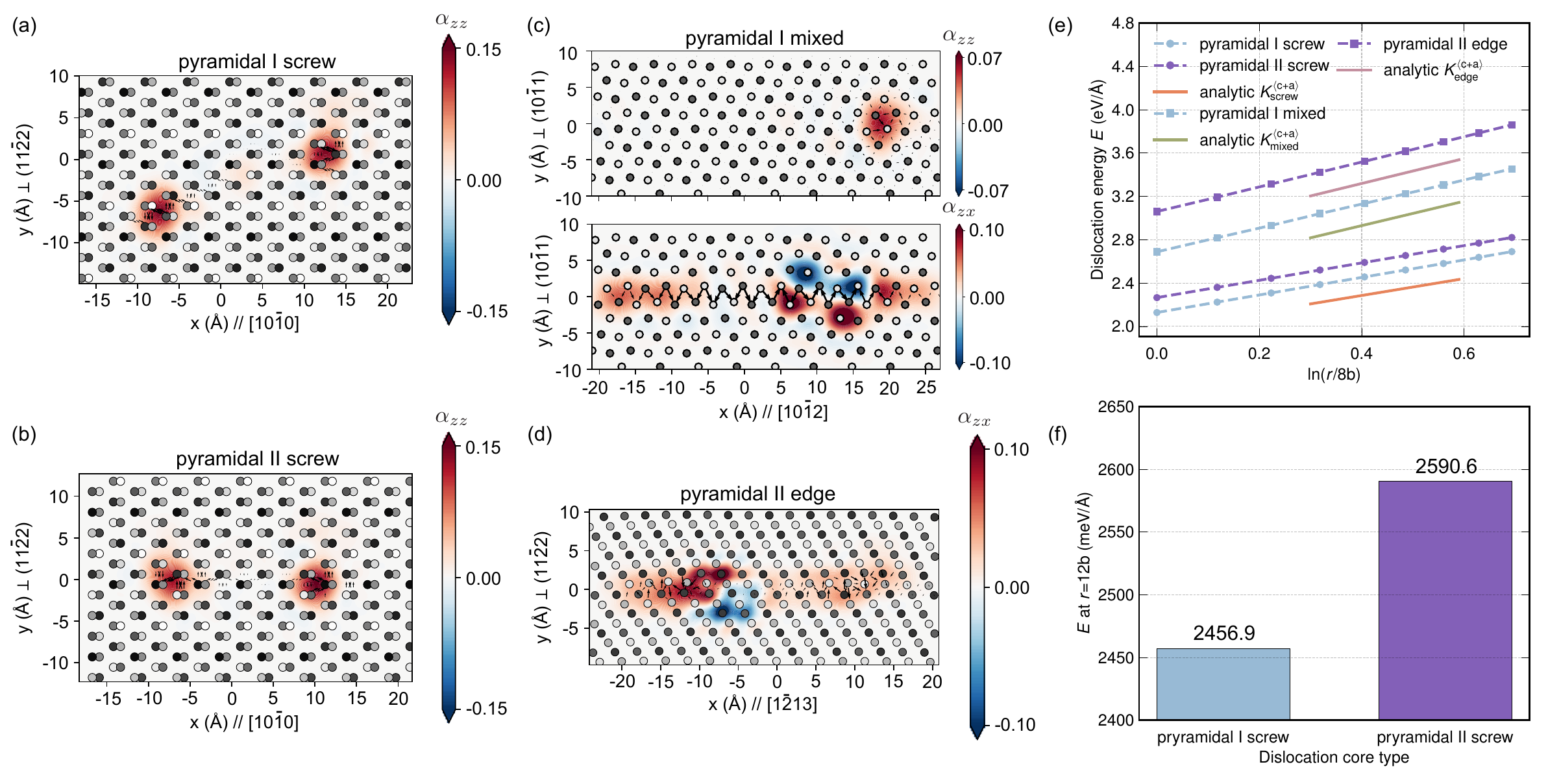}
	\caption{\label{fig:ca_hcp} \cadisl dislocation core structures and energies in HCP Ti. (a-b) Screw \cadisl dislocation on the pyramidal I and II planes. (c) Mixed \cadisl dislocation core structure on the pyramidal I plane. $\alpha_{zz}$ and $\alpha_{zx}$ show the screw and edge components, respectively.  (d) Edge \cadisl dislocation on the pyramidal II plane. All cores are visualized by the Nye tensor ($\alpha_{zz}$ and $\alpha_{zx}$, \AA\(^{-1}\)) and differential displacement plots. (e) Energy of the \cadisl dislocations in (a-d) as a function of \(\ln{(r/8b)}\).  The slopes \(K_{\rm screw/mixed/edge}\) are calculated using anisotropic linear elasticity theory~\cite{hirth_1992_disl}. (f) Energies of the screw \cadisl dislocations on pyramidal I and II planes at \(r = 12b\).}
\end{figure*}

Figure~\ref{fig:ca_hcp} shows the core structures of screw, mixed and edge \cadisl dislocations predicted by DP-Ti. The screw dislocation can dissociate into a pair of partials with a stacking fault on either the pyramidal I or II plane, depending on the initial core construction.  When a compact/non-dissociated core is introduced, it always dissociates onto the pyramidal I plane (Fig.~\ref{fig:ca_hcp}a), similar to that in DFT (Fig.~\ref{fig:ca_screw_hcp}a-d).  The pyramidal II dissociation (Fig.~\ref{fig:ca_hcp}b) can be obtained by creating a pair of partials of \hcadisl separated by \(\sim\)10 \AA\ on the pyramidal II plane.  Both partials have nearly pure screw character and the partial cores are similar, independent of the dissociation plane (Figs.~\ref{fig:ca_hcp}a, b).   The mixed \cadisl dislocation dissociates into two partials on the pyramidal I plane (Fig.~\ref{fig:ca_hcp}c).  The partial core on the right has a mixed character with a screw component of \hadisl aligned on a second pyramidal I plane, resulting in a non-planar partial core, while the left partial core is of pure edge character. The screw component of the right partial is similar to that of a split/saddle core in BCC transition metals (cf. Ref.~\cite{dezerald_2014_prb} and Fig.~\ref{fig:ca_hcp}c).  This non-planar dissociation may lead to large barriers to the glide of the mixed \cadisl dislocations.  Finally, the edge dislocation adopts a planar dissociation on the pyramidal II plane; both of its partials are of pure edge character, similar to that of Mg~\cite{wu_2015_msmse}.  In all cases, the elastic energy pre-factors \(K\) measured from atomistic simulations agree very well with the anisotropic elasticity analytical predictions (Table~\ref{tab:elastic_k}).  For the screw \cadislns, the pyramidal II dissociation has a higher total/near-core energy \(\Delta E\) of 134 meV/\AA; this is nearly an order of magnitude larger than that for the screw \adisl dislocation on the pyramidal I and prism I planes.  The pyramidal II \cadisl screw is thus metastable; additional simulations show that it spontaneously cross-slips to the pyramidal I plane at 4 K.  This is consistent with experimental observations where \cadisl slip is dominant on the pyramidal I plane in pure Ti~\cite{numakura_1986_sm,paton_1970_mt,pochettino_1992_sm,zaefferer_2003_msea,gong_2009_am,wang_2013_mmta,kishida_2020_am}.

\begin{figure}[!htbp]
	\includegraphics[width=1.0\textwidth]{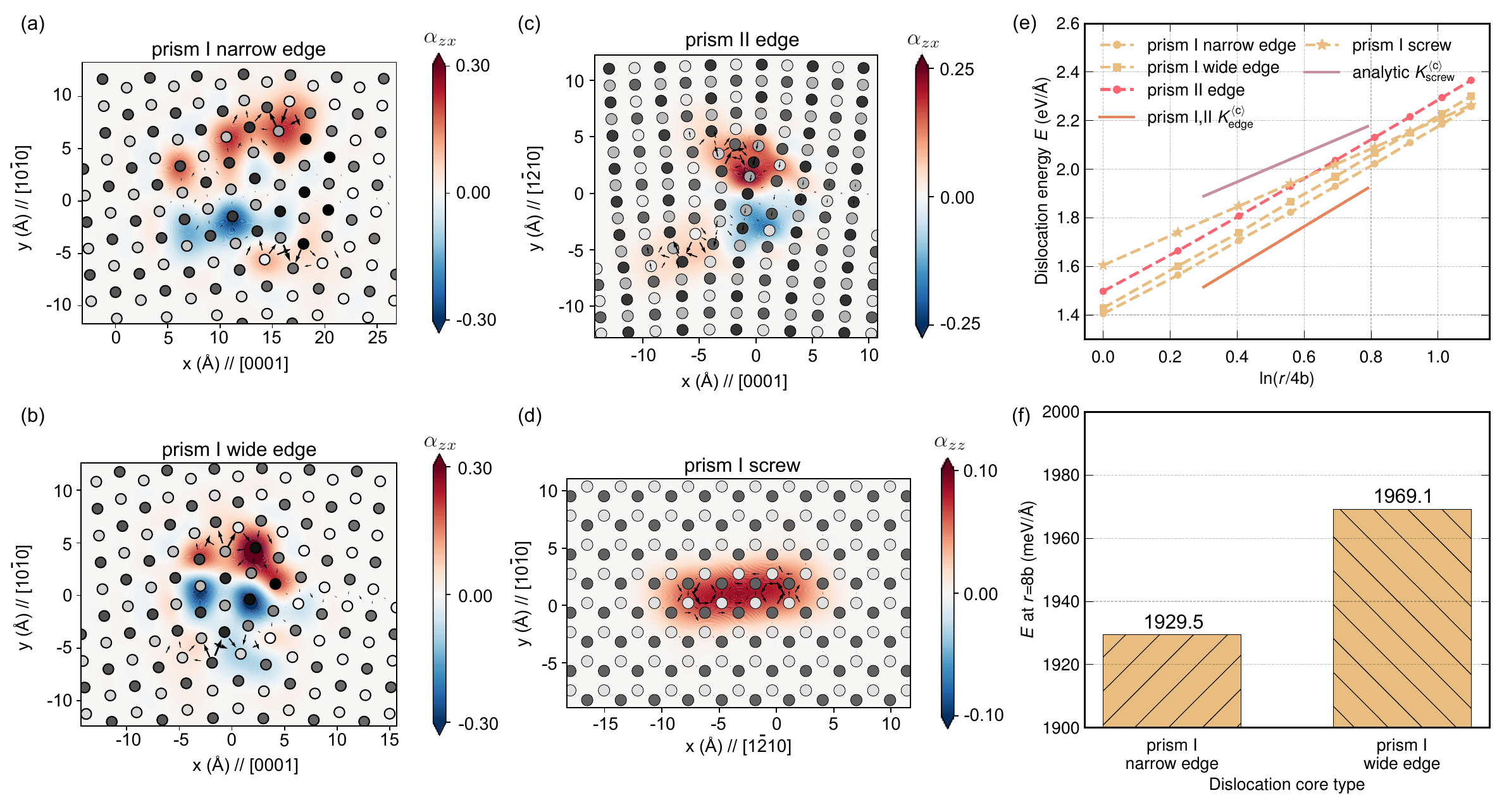}
	\caption{\label{fig:c_hcp} \cdisl dislocation core structures and energies in HCP Ti. (a-d) Edge and screw \cdisl dislocation core structures visualized by the Nye tensor ($\alpha_{zx}$ and $\alpha_{zz}$, \AA\(^{-1}\)) and differential displacement plots. (e) Energy of edge and screw \cdisl dislocations calculated from Eq.~\eqref{eq:disl_energy}. (f) Energies of the edge \cdisl dislocations on prism I narrow and wide planes at \(r=8b\). In (a-c), the climb-dissociated core may adopt slightly different structures depending on the initial core position relative to the atomic lattice. }
\end{figure}

Figure~\ref{fig:c_hcp} shows the core structures and energies of the \cdisl dislocations. The edge dislocation can be constructed along three planes: the prism I narrow, prism I wide and prism II planes.  In all the cases, the edge core dissociates onto the basal plane, which is perpendicular to the respective glide prism planes (Figs.~\ref{fig:c_hcp}a-c).  This dissociation requires the two partial cores to climb upwards and downwards relative to the initial glide plane, leading to sessile cores (as seen in HCP Mg~\cite{wu_2015_nature}). The dissociation creates an I\(_{\rm 1}\) stacking fault on the prism I narrow, wide and prism II planes.

The screw \cdisl dislocation always dissociates into a pair of partials with a stacking fault on the prism I wide plane.  The two partials are of mixed character with edge components of \(\pm\)\adislns/2 (Fig.~\ref{fig:gamma_surface_hcp}). The associated stacking fault energy is 0.697 J/m$^2$ in DP-Ti; lower than the DFT value of 1.034 J/m\(^2\)~\cite{yin_2017_am}.  The planar dissociation width is $\sim$2$a$, larger than the prediction based on elasticity.  The prism II plane has a metastable stacking fault of energy 0.515 J/m$^2$, lower than that on the prism I plane. However, the screw core always prefers dissociation onto prism I planes; the prism II dissociation at the metastable stacking fault position creates edge components of \(\sim\)0.8\adisl and is thus energetically unfavourable relative to the prism I dissociation.

For the edge \cdislns, the energy pre-factor \(K_{\rm edge}^{\langle  \mathbf{c} \rangle}\) on the prism I plane is the same as that of dislocation on the prism II plane based on the analytic anisotropic elasticity and MD simulations (Table~\ref{tab:elastic_k}). Dissociation on the prism I narrow plane has lower near-core energy than dissociation on the prism I wide plane.  However, all climb dissociations are sessile and not in thermodynamic equilibrium; further climb dissociations are expected at higher temperatures or longer time scales where local diffusion is sufficiently activated.  Since the screw \cdisl always favours the prism I wide plane, slip via the \cdisl dislocation is expected to occur primarily on this plane.  Nevertheless, since the edge dislocation is not stable on the slip plane, \cdisl slip can not provide strain accommodation easily in HCP Ti. In all cases, the energy pre-factors of the \cdisl edge are always larger than their screw counterpart (Table~\ref{tab:elastic_k}).

\subsection{\label{sec:hcp_mobility}Critical Resolved Shear Stress and Dislocation Mobility}

The above study examines all the possible dislocations in HCP Ti.  Overall, DP-Ti reproduces the various dislocation cores largely consistent with previous and current DFT calculations. DP also enables us to explore the widely dissociated cores which are impractical to study by DFT.  The dissociation planes are also largely consistent with the primary slip systems of HCP Ti reported in various experiments.  We now examine the glide behaviour of the dislocations under driving stresses, both at 0 K and 300 K.  We focus on the glissile cores which dissociate on the respective slip planes.

\begin{table}[!htbp]
	\centering
  \begin{threeparttable}[b]
	\small
	\caption{\label{tab:crss_a_hcp} Critical resolved shear stress ($\tau_0$ in MPa) of \adisl dislocations measured using DP-Ti and the experimental data.}
		\begin{tabular}{p{3cm}p{1cm}p{2cm}p{1cm}p{1cm}p{1.5cm}p{2cm}}
			\hline
			&\multicolumn{2}{c}{\textbf{Basal plane}} & \multicolumn{2}{c}{\textbf{Prism I wide plane}} & \multicolumn{2}{c}{\textbf{ Pyramidal I narrow plane}} \\
      \hline
			Model and temperature &Edge&Screw&Edge&Screw&Edge&Screw\\
			DP-Ti (0 K) & 151 &\(\sim\)+1600-1650\tnote{1} & 277 & 111 & 280 &$>$600\tnote{5}\\
			DP-Ti (300 K)& $<$5 & \(<\)5 &$<$5&$<$5&$<$5 & $\sim$10-20 MPa\tnote{6}\\
			Exp ($\sim$300 K)& \multicolumn{2}{c}{85\tnote{2}, 209\tnote{3}}  & \multicolumn{2}{c}{20-50\tnote{4}, 181\tnote{3}} &\multicolumn{2}{c}{-}\\
			\hline
		\end{tabular}
		\begin{tablenotes}
		  \footnotesize
      \item [1] The dislocation glides on the pyramidal I plane at high shear stresses on the basal plane.
		  \item [2] Simple shear tests of bulk single-crystals of commercially pure HCP-Ti (CP-Ti)~\cite{levine_1966_tmsaime}.
      \item [3] Self-consistent crystal plasticity finite element modelling fitted to data bending of single crystals CP-Ti micro-cantilever~\cite{gong_2009_am}.
		  \item [4] Uni-axial tension, uni-axial compression, and simple shear tests of bulk single crystals of CP-Ti~\cite{levine_1966_tmsaime,paton_1970_mt,cass_1970_stat,tanaka_1972_am,tanaka_1971_2_am,akhtar_1975_mmta}.
		  \item [5] The pyramidal core cross-slips to and glides on the prism I plane.
		  \item [6] The pyramidal core cross-slips to and glides on the prism I plane at \(\tau= \) 20 MPa after 520 ps in MD (Fig.~\ref{fig:a_lock_unlock_hcp}b).
		\end{tablenotes}
	\end{threeparttable}
\end{table}

Table~\ref{tab:crss_a_hcp} shows the CRSS ($\tau_0$) of the \adisl dislocations on different slip planes measured using DP-Ti and compared with experimental values from bulk single-crystals of commercially pure HCP-Ti.  The screw \adisl is unstable on the basal plane in both DFT~\cite{kwasniak_2019_sm} and DP-Ti calculations.  We thus measure the response of the screw \adisl dissociated on the pyramidal I plane with a shear stress applied on the basal plane.  In experiments, the CRSS~\cite{levine_1966_tmsaime} can not be unambiguously assigned to individual dislocation of specific character.  At 0 K, the screw dislocation is stationary when the shear stress is applied on the basal plane for stress below 1600 MPa, above which it glides on the pyramidal I plane.  The screw \adisl on the prism I wide plane has the lowest CRSS compared to those on basal and pyramidal I planes.  This is consistent with the screw partial core structures and relative energetics reported above.  Specifically, when the screw \adisl glides on the prism I wide plane, its partial cores always pass through similar configurations/local atomic environments with similar energies. In contrast, when the screw \adisl glides on the pyramidal I narrow plane, two metastable partial cores pass through configurations with distinct differences in structures and energies.  This is similar to the saddle and easy cores in BCC structures with large differences in structures and energies. Nonetheless, the saddle core in BCC may be unstable while both cores are at least meta-stable on the pyramidal I narrow plane in HCP Ti. In all cases, the CRSS values are sequenced as follows: \(\tau_{\rm basal }^{\rm screw} > \tau_{\rm pyramidal\ I }^{\rm screw} > \tau_{\rm pyramidal\ I }^{\rm edge} > \tau_{\rm prism\ I }^{\rm edge} > \tau_{\rm basal }^{\rm edge}  > \tau_{\rm prism\ I }^{\rm screw}  \).

\begin{figure*}[!htbp]
	\includegraphics[width=1.0\textwidth]{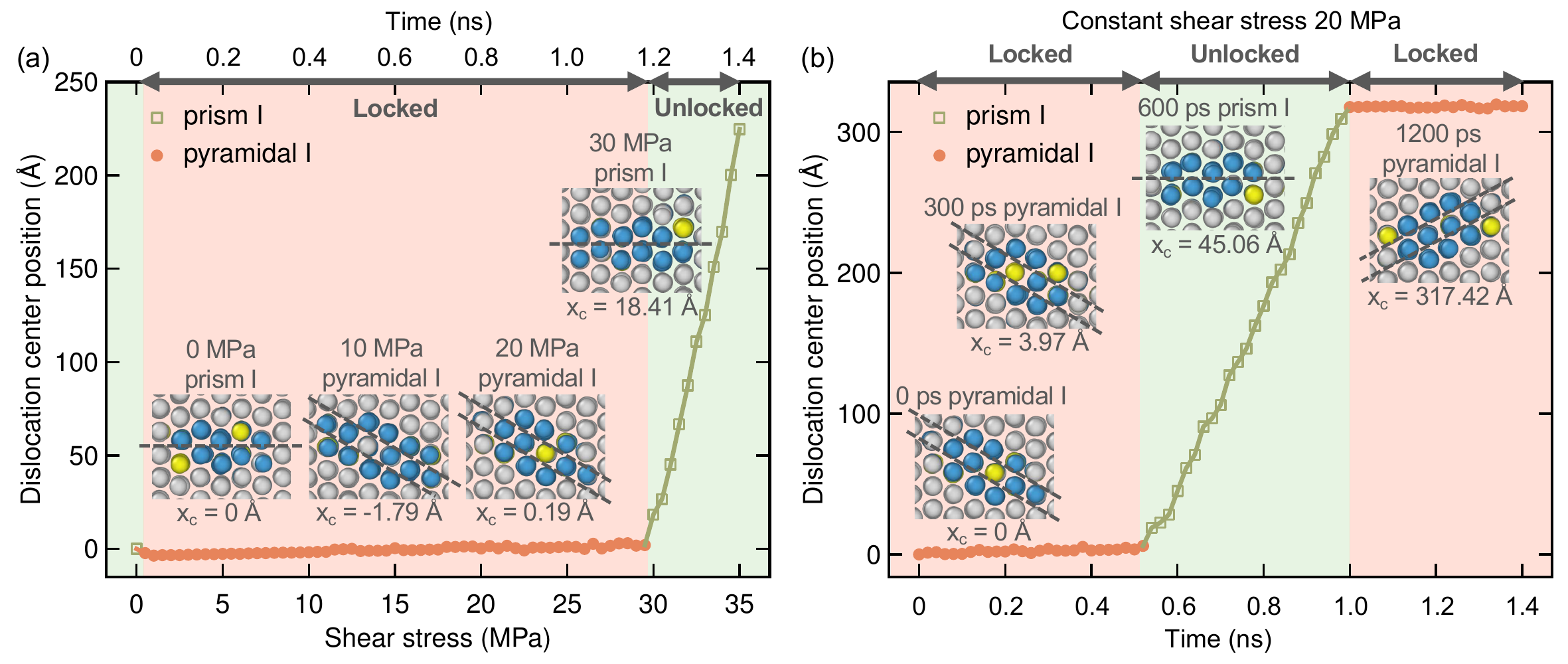}
	\caption{\label{fig:a_lock_unlock_hcp} ``Locking-unlocking" core transition during the stress ramping process and under constant shear stress. \(x_c\) is the position of the dislocation centre. (a) The relationship between dislocation centre position and applied shear stress. (b) The relationship between dislocation centre position and simulation time at a constant shear stress of 20 MPa. The insets show different core structures under different stresses or at different simulation times.  Atoms are coloured based on their local atomic environment/structure as identified by the common neighbour analysis (CNA~\cite{faken_1994_cms}); HCP: grey, BCC: yellow, others: blue.}
\end{figure*}

The \adisl dislocations are highly mobile at 300 K, except for the screw core on the pyramidal I plane.  Their CRSSs are negligible compared with those at 0 K.  The CRSS values are difficult to extrapolate from fitting to Eq.~\eqref{eq:mobility} and should be viewed accurate within a few MPa.  In contrast, the screw core appears to be sessile on the pyramidal I plane when the applied shear stress is below 10 MPa.  At higher applied stresses and with longer waiting time, it exhibits a ``locking-unlocking" behaviour similar to that seen in \textit{in-situ} experiments~\cite{clouet_2015_natmat}.  To further investigate this phenomenon, we perform MD simulations at a lower stress-ramp rate of 25 MPa/ns at 300 K.  We use a dislocation of \(\sim\)6 nm in length so that the simulation time can be extended to ns time scales.  The screw \adisl dislocation is constructed on the prism I wide plane, which quickly transforms onto the pyramidal I narrow plane at a small applied shear stress (\(\sim\)0.5 MPa).  Figure~\ref{fig:a_lock_unlock_hcp}a shows the dislocation position and core structure during the stress-ramp process. The dislocation is nearly stationary in the locked configuration for the first 1.2 ns. The pyramidal I core cross-slips to the prism I plane when \(\tau\) reaches \(\sim\)30 MPa. The cross-slipped core then glides quickly on the prism I plane in the unlocked configuration for stress from 30 to 35 MPa during the stress ramp.

To investigate the core transition in detail, we perform MD simulations at a constant applied shear stress of 20 MPa under a constant temperature of 300 K.  The dislocation core is initially dissociated on the pyramidal I plane, the same as that in Fig.~\ref{fig:a_lock_unlock_hcp}a. The entire dislocation is nearly stationary with one of its partial core wandering to the next available core position (Fig.~\ref{fig:screw_a_hcp}). The effective velocity is \(1.3 \times 10^{-5}\) {\AA}/ps.  The pyramidal I dissociated structure thus corresponds to the ``locked'' configuration of the screw \adisl dislocation.  At \(t \approx520\) ps (520000 time steps), the pyramidal I core cross-slips to the prism I plane (Fig.~\ref{fig:a_lock_unlock_hcp}b) and starts gliding at a nearly constant speed of 0.6 {\AA}/ps, which corresponds to the unlocked core configuration.  At \(t \approx 988\) ps, the gliding dislocation cross-slips back to the pyramidal I plane and becomes stationary again, i.e., transforms to the ``locked'' core.  The current DP-Ti thus exhibits ``locking-unlocking'' screw \adisl core behaviour in pure Ti (without oxygen or other interstitial effects).

In the \textit{in-situ} TEM straining experiments~\cite{clouet_2015_natmat}, the screw \adisl dislocation appeared locked on the pyramidal I plane for several seconds, followed by a quick glide on the prism I plane.  At least two factors may contribute to the discrepancy in time scales between experiments and the MD simulations.  First, the temperatures are different in the experiment and MD simulations: the experiment was conducted at 150 K, while the MD simulation is performed at 300 K.  MD simulations are also performed at 150 K but no ``locking-unlocking" transition is observed within 2 ns. This indicates that the waiting time required for core transition at 150 K is beyond the time scale accessible by current MD simulations.  The higher temperature in MD simulations should in principle make the core transform occur more easily.  Second, the sample sizes are different in the simulations and in the experiment.  Although we have used a large supercell with over 0.15 million atoms and a dislocation line of $\sim$6 nm,  this supercell is still much smaller than the experimental sample.  Dislocation core transitions occur stochastically and longer dislocation length may favour the transition to shorter waiting times since each segment can act as independent nucleation sites for the transition.  The waiting and gliding times may not be quantitatively comparable between the simulations and experiments.  Nevertheless, this ``locking-unlocking'' mechanism in MD should be similar to that in the \textit{in-situ} TEM~\cite{clouet_2015_natmat} and is an intrinsic behaviour of the screw \adisl dislocation in Ti.

\begin{figure}[!htbp]
	\centering
	\includegraphics[width=1.0\textwidth]{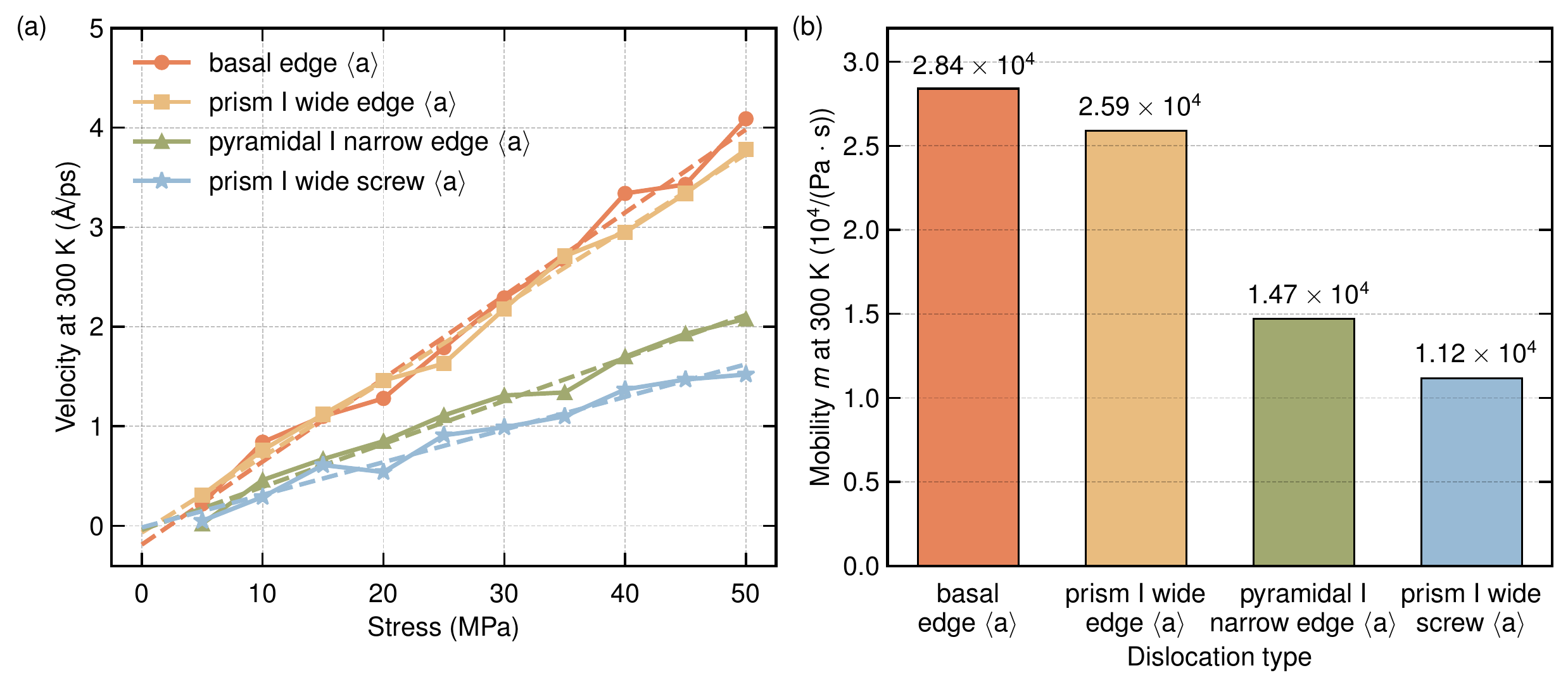}
	\caption{\label{fig:mobility_a_hcp} \adisl dislocation velocities and mobilities measured from MD simulations at 300 K. (a) Dislocation velocities as a function of applied shear stress. (b) Dislocation mobilities $m$ fitted according to Eq.~\eqref{eq:mobility}.}
\end{figure}

Figure~\ref{fig:mobility_a_hcp} shows the velocities and mobilities of the \adisl dislocations measured in MD simulations at 300 K.  On the pyramidal I plane, the screw \adisl dislocation mobility cannot be unequivocally determined as it cross-slips to the prism I plane at high stresses (Fig.~\ref{fig:a_lock_unlock_hcp}a).  The screw dislocation is not stable on the basal plane, as discussed earlier.  For the remaining cases, the velocity increases nearly linearly with increasing applied stresses, allowing the mobility to be fitted robustly using Eq.~\eqref{eq:mobility}.  The edge \adisl dislocation on the basal plane has the highest mobility, followed closely by the edge \adisl dislocation on the prism I wide plane.  The mobility of the edge dislocation on the pyramidal I narrow plane is nearly half of the previous two cases, while the screw \adisl on the prism I wide plane has the lowest mobility.  Among all the cases, the dislocation mobilities are of the same order of magnitude and the differences are less drastic compared to those in BCC transition metals~\cite{gilbert_2011_prb,queyreau_2011_prb}.

\begin{figure}[!htbp]
  \centering
	\includegraphics[width=0.9\textwidth]{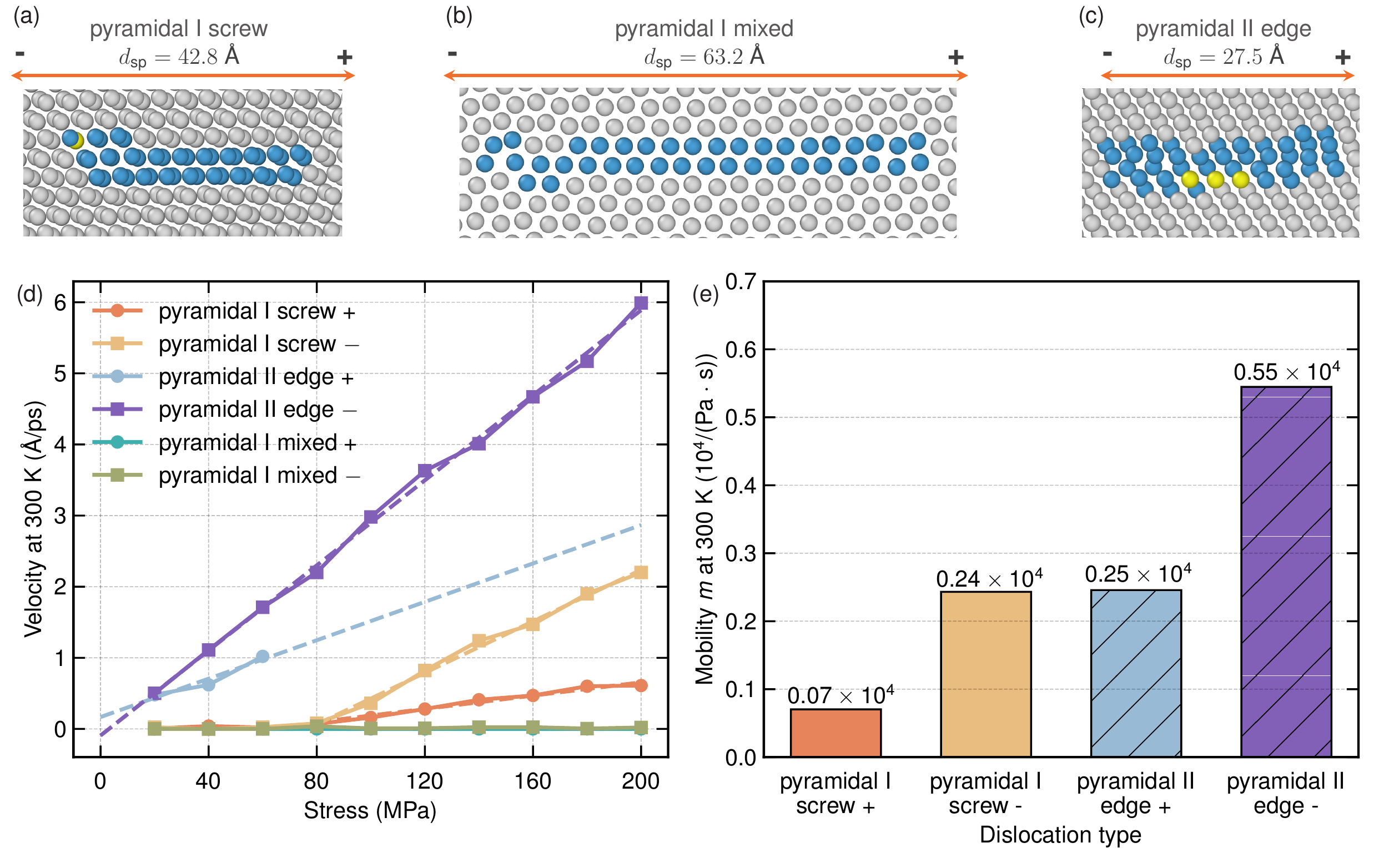}
	\caption{\label{fig:mobility_ca_hcp} \cadisl dislocation velocities and mobilities calculated by DP-Ti at 300 K. (a-c) Conventions of the positive and negative shear. \(d_\text{sp}\) is the partial dislocation separation distance calculated by anisotropic linear elasticity theory from~\cite{yin_2017_am}. See Fig.~\ref{fig:a_lock_unlock_hcp} for the atom colour scheme. (d) Dislocation velocity as a function of applied shear stress. (e) Mobility $m$ from fitting the simulation data according to Eq.~\eqref{eq:mobility}.}
\end{figure}

Table~\ref{tab:crss_ca_disl} shows the CRSS of \cadisl dislocations calculated by DP-Ti.  The Burgers vector of the two partials are different, so are their core structures and responses to applied stresses.  In particular, the glide of the dislocation has directionality~\cite{wu_2015_nature}.  Figure~\ref{fig:mobility_ca_hcp} shows the positive/negative direction convention used here. At 0 K, the mixed \cadisl exhibits very high CRSSs in both directions.  The screw \cadisl can glide smoothly in both directions on the pyramidal I plane; the CRSSs differ by 221 MPa between the two directions.  In contrast, the \cadisl screw dislocation is unstable on the pyramidal II plane under applied shear stresses, while the \cadisl edge can glide in both directions with a CRSS difference of 16 MPa.  At 300 K, the applied shear stress of 200 MPa can hardly drive the mixed \cadisl to move in the positive and negative directions. The CRSSs of the screw decrease by nearly an order of magnitude and the directionality is substantially reduced.

\begin{table}[!htbp]
	\centering
	\begin{threeparttable}[b]
		\small
		\caption{\label{tab:crss_ca_disl} Critical resolved shear stress ($\tau_0$ in MPa) of \cadisl dislocation measured from experiments and calculated by DP-Ti.}
			\begin{tabular}{p{4cm}p{2.5cm}p{2cm}p{2cm}p{2cm}}
				\hline
				&\multicolumn{2}{c}{\textbf{Pyramidal I plane}} &\multicolumn{2}{c}{\textbf{Pyramidal II plane}}\\
				Model and temperature &Mixed&Screw&Edge&Screw\\
        \hline
				\multirow{2}{*}{DP-Ti (0 K)} & \(\sim\)+1200-1400 &+621&+523& \multirow{2}{*}{not stable} \\
				&-1117&-842&-507&\\
				\multirow{2}{*}{DP-Ti (300 K)}&$>$+200&+60& \multirow{2}{*}{not stable\tnote{4}} & \multirow{2}{*}{not stable}\\
				&$\sim$-200&-76&&\\
				Exp ($\sim$300 K)& \multicolumn{2}{c}{474\tnote{1}, 580-635\tnote{2}, 765\tnote{3}} &\multicolumn{2}{c}{-}\\
				\hline
			\end{tabular}
			\begin{tablenotes}
				\footnotesize
				\item [1] Self-consistent crystal plasticity finite element modelling fitted to data bending of single crystals CP-Ti micro-cantilever~\cite{gong_2009_am}.
				\item [2] Uni-axial compression of [0001]-oriented single crystal CP-Ti micro-pillar~\cite{kishida_2020_am}.
				\item [3] Uni-axial compression of [0001]-oriented single crystal Ti-5Al micro-pillar~\cite{yu_2013_sm}.
				\item [4] When edge \cadisl moves in the positive direction at 300 K, it will dissociate on basal plane at the shear stress higher than 65 MPa, similar to that in Fig.~\ref{fig:ca_hcp_highT}.
			\end{tablenotes}
		\end{threeparttable}
\end{table}

Figure~\ref{fig:mobility_ca_hcp} shows the mobilities of the \cadisl dislocations. The edge dislocation is highly mobile on the pyramidal II plane; it has the highest mobility in the negative direction while its mobility is reduced by 50\% in the positive direction. Furthermore, the edge is not stable at 300 K and dissociates onto the basal plane when the applied shear stress is above 65 MPa, similar to that observed in Mg~\cite{wu_2015_nature} and discussed below for high temperature behaviour.  The screw \cadisl exhibits a distinct threshold stress (taken as the CRSS \(\tau_0\)) below which the dislocation is nearly stationary. Above \(\tau_0\), the velocity is proportional to the net shear stress \(\tau-\tau_0\).  Its mobility also has strong directionality; the value in the negative direction is nearly 3.5 times of that in the positive direction (Fig.~\ref{fig:mobility_ca_hcp}e). The mixed dislocation has a very low velocity (\(< 4 \times 10^{-6}\) {\AA}/ps) and the mobility is almost 0 for applied shear stresses $<$ 200 MPa. In all the cases, the edge \cadisl has the highest mobility on the pyramidal II plane, followed by the screw \cadisl and mixed \cadisl on the pyramidal I plane.

The above results show that the edge (under positive shear) and screw (under both positive and negative shears) \cadisl dislocations are unstable on the pyramidal II plane at room temperature. It is thus difficult to sustain \cadisl slip on the pyramidal II plane in HCP Ti, leaving the pyramidal I plane as the primary slip plane for \cadisl dislocations, which is consistent with a wide range of experimental observations~\cite{numakura_1986_sm,paton_1970_mt,pochettino_1992_sm,zaefferer_2003_msea,gong_2009_am,wang_2013_mmta,kishida_2020_am}. Furthermore, on the pyramidal I plane, the mixed \cadisl shows a much lower mobility relative to the pure screw \cadislns.  Consequently, the mixed \cadisl should be the primary dislocations observed in experiments - also consistent with a previous TEM study~\cite{numakura_1986_sm} and a recent \textit{in-situ} TEM observation where \cadisl dislocations bow out from a tangle of \adisl dislocations and some straight \cadisl segments are stationary or pinned~\cite{yu_2013_sm}.

\begin{figure}[!htbp]
  \centering
	\includegraphics[width=0.9\textwidth]{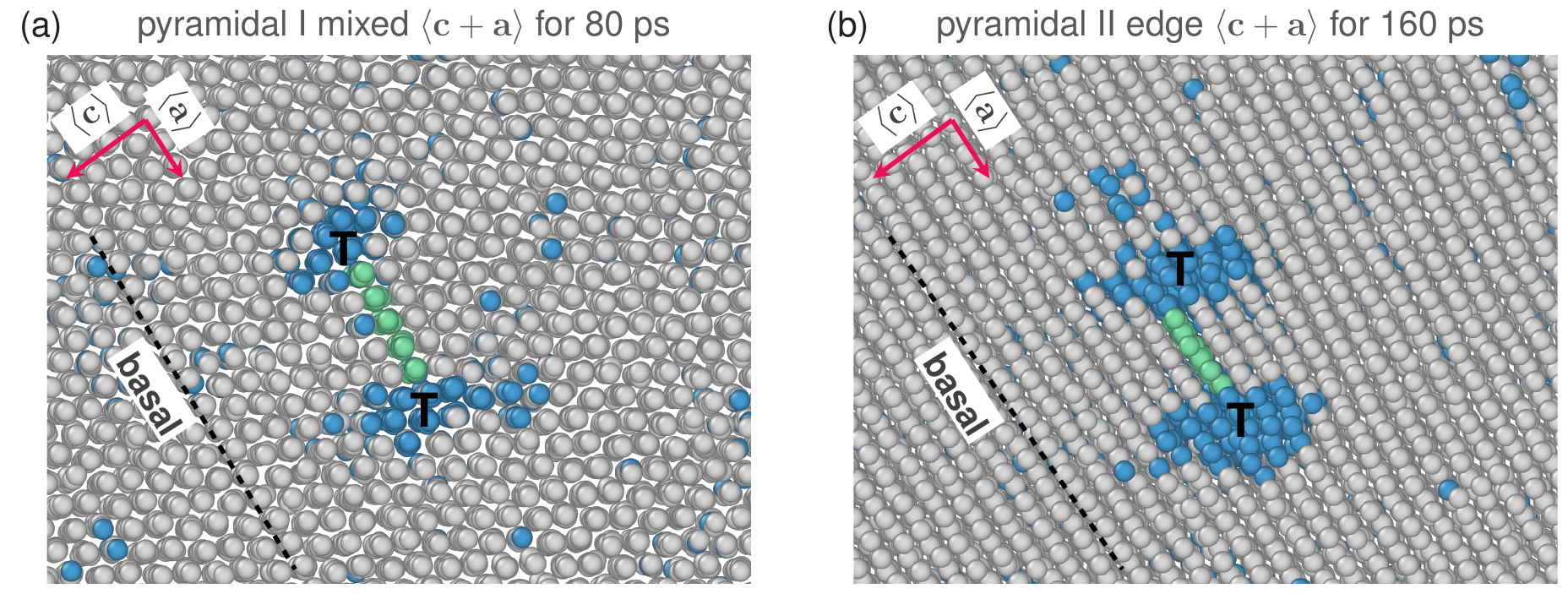}
	\caption{\label{fig:ca_hcp_highT} Basal-dissociated dislocations at 900 K. (a) Mixed \cadisl dislocation climb-dissociated along the basal plane from the pyramidal I plane after 80 ps in MD simulations. (b) Edge \cadisl dislocation climb-dissociated on the basal plane from the pyramidal II plane after 160 ps in MD simulations. See Fig.~\ref{fig:a_lock_unlock_hcp} for the atom colour scheme. }
\end{figure}

Finally, we examine the high temperature stability of the \cadisl mixed and edge dislocations on the pyramidal I and II planes.  An earlier dislocation energetic analysis~\cite{wu_2016_am} suggests that \cadisl dislocations are not stable against climb dissociations on the basal plane in all HCP structures.  Figure~\ref{fig:ca_hcp_highT} shows the two dislocations obtained in MD simulations at 900 K.  In both cases, the dislocations are first constructed on their respective pyramidal planes and are stable for \(\sim\)30 ps and \(\sim\)40 ps.  The two dislocations then spontaneously undergo a climb-dissociation to the final basal-dissociated configurations with no stresses applied in the simulations.  This climb-dissociation is similar to that seen in HCP Mg~\cite{wu_2015_nature,wu_2016_sm}.  The climb-dissociated cores have lower energies than their pyramidal-dissociated counterparts.  In the climb-dissociated configurations,  the two partials reside on different pyramidal planes and are connected by an \(\text{I}_\text{1}\) stacking fault on the basal plane.  The transformed \cadisl dislocation is sessile, which reduces the mobile \cadisl dislocation density and provides barriers to other \cadisl dislocations on the pyramidal planes.  This intrinsic transformation and the presence of sessile cores inhibit further \cadisl dislocation slip and may explain the anomalous shear instability of HCP Ti when compressed along the \cdisl axis at high temperatures~\cite{williams_2002_mmta}.

\subsection{\label{sec:hcp_twin_energy}Twin Boundary Structures and Energies}

\begin{figure}[!htbp]
	\centering
	\includegraphics[width=1.0\textwidth]{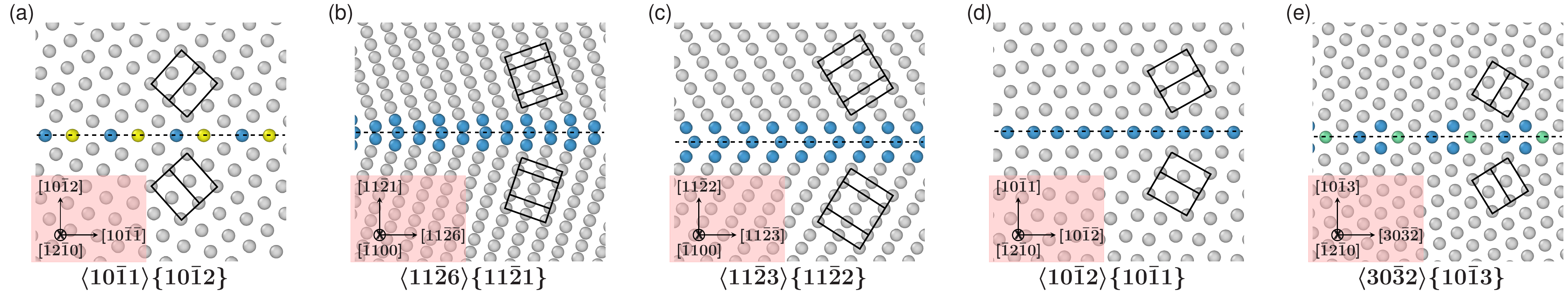}
	\caption{\label{fig:twin_hcp} Twin boundary structures relaxed with DP-Ti. See Fig.~\ref{fig:a_lock_unlock_hcp} for the atom colour scheme. }
\end{figure}

Deformation twinning is one of the primary deformation mechanisms in HCP Ti.  It becomes prevalent with decreasing temperature~\cite{zhao_2021_science}.  Figure~\ref{fig:twin_hcp} shows 5 optimised, coherent twin boundary (TB) structures predicted by DP-Ti.  Table~\ref{tab:hcp_twin_energy} shows their energies in comparison with DFT results~\cite{kumar_2015_am,hooshmand_2017_msmse}.  The first 4 TBs are commonly observed in experiments, while the last one on the \(\{10\bar{1}3\}\) plane was reported recently~\cite{kumar_2015_am,hooshmand_2017_msmse}.  In all cases, the interatomic potential DP-Ti reproduces the TB energies within 10\% deviation from the corresponding DFT values.  DP-Ti thus reproduces all TB energies and are appropriate for MD studies of deformation twinning and dislocation-twin interactions in HCP Ti.

\begin{table}[!htbp]
	\centering
  \begin{threeparttable}[b]
	\small
	\caption{\label{tab:hcp_twin_energy} Twin boundary energies calculated by DFT and DP-Ti.}
		\begin{tabular}{p{3cm}p{1.8cm}p{1.8cm}p{2.2cm}p{1.8cm}p{1.8cm}}
			\hline
			&\multicolumn{5}{c}{Twin boundary energy (mJ/m$^2$)}\\
			& \{10$\bar{1}$2\} &\{11$\bar{2}$1\} & \(\{11\bar{2}2\}\) &\{10$\bar{1}$1\} & \{10$\bar{1}$3\}  \\
			\hline
			DFT&297.6\cite{kumar_2015_am}, 308.4\cite{hooshmand_2017_msmse}, 330\cite{hooshmand_2021_arxiv} &233.2\cite{lane_2011_prb}, 273.1\cite{hooshmand_2021_arxiv} & 377.1(this work), 385\cite{hooshmand_2021_arxiv} & 75.3\cite{kumar_2015_am}, 100.3\cite{hooshmand_2017_msmse}, 119.4\cite{hooshmand_2021_arxiv} &326.5\cite{kumar_2015_am}, 345.1\cite{hooshmand_2017_msmse}\\
			DP-Ti&276.4&226.8 & 350.1 
                                           &126.5&318.4\\
      \hline
		\end{tabular}
	\end{threeparttable}
\end{table}

\section{\label{sec:bcc} Dislocations in BCC-\(\beta\) Ti}

\subsection{\label{sec:bcc_core} Dislocation Core Structures}

\begin{figure}[!htbp]
	\centering
	\includegraphics[width=1.0\textwidth]{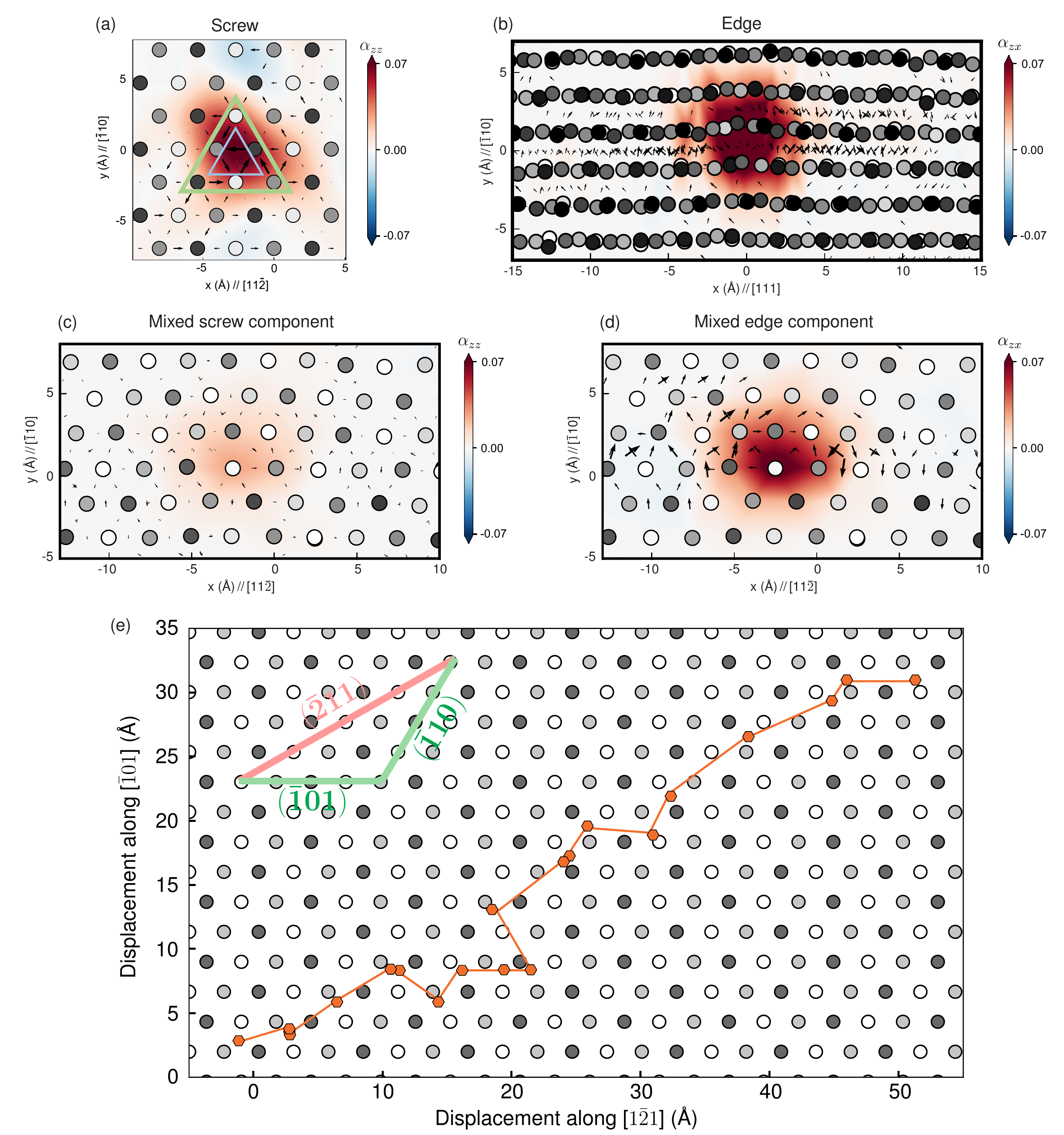}
	\caption{\label{fig:disl_a_bcc} Core structures of the $\langle 111 \rangle/2$ dislocations at 1000 K and $\langle 111 \rangle/2$ screw dislocation glide behaviour at 1400 K obtained in MD simulations using DP-Ti. (a) Degenerate core structure of the screw dislocation. (b) Edge dislocation core. (c-d) Screw and edge component of 70.5\(^\circ\) mixed dislocation. The atoms are plotted based on their positions averaged in 5 ps. $\alpha_{zz}$ (\AA$^{-1}$) and $\alpha_{zx}$ (\AA$^{-1}$) show the screw and edge components, respectively. In (b), the edge dislocation wanders on the slip plane during the measurement, resulting in distorted atomic columns in the plot.  (e) Trajectory of the screw dislocation at an applied shear stress of 160 MPa for 40 ps at 1400 K. The time interval between two neighbouring points is 2 ps.}
\end{figure}

\begin{figure*}[!htbp]
	\includegraphics[width=1.0\textwidth]{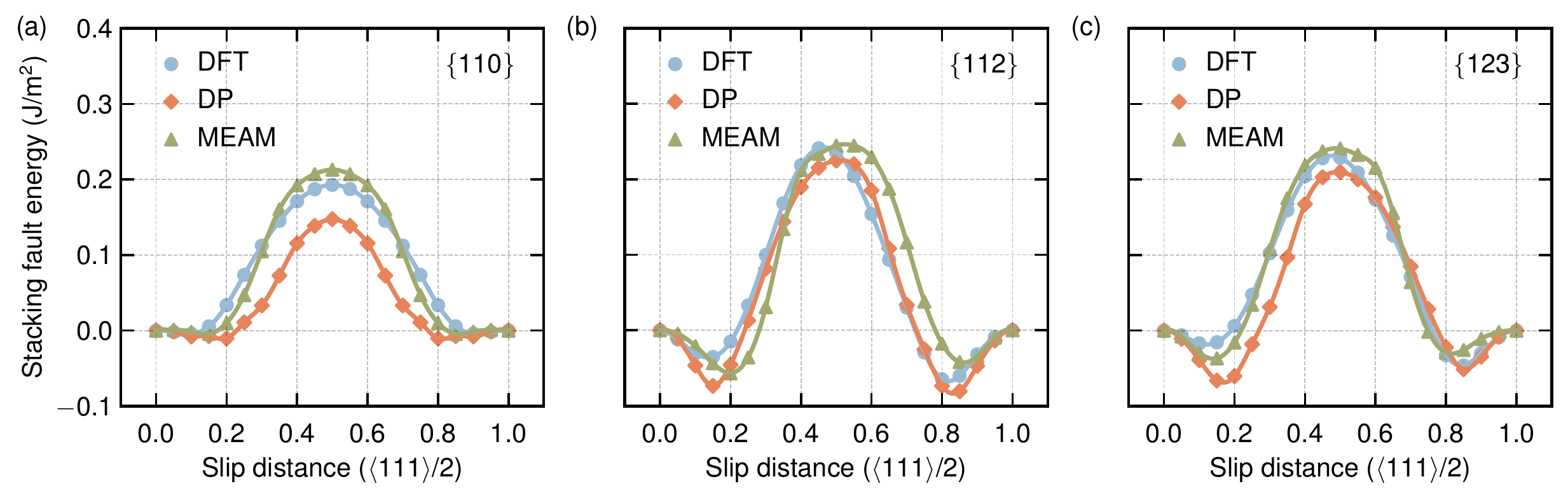}
	\caption{\label{fig:gamma_line_bcc} The generalised stacking fault energy ($\gamma$) lines of BCC Ti calculated by DFT, DP-Ti, and an MEAM potential~\cite{hennig_2008_prb}.}
\end{figure*}

HCP-\(\alpha\) Ti is stable at moderate temperatures and transforms to BCC-\(\beta\) Ti at 1155 K under zero pressure conditions~\cite{tonkov_2005_crc}. The BCC structure is neither stable nor metastable at 0 K.  It is thus challenging to study the dislocation core properties in BCC-Ti in first-principles DFT calculations or experiments.  The current DP-Ti predicts a HCP to BCC phase transition at 1252 K under zero stress, similar to that in experiment (Fig.~\ref{fig:ti_phase_diagram}).  In addition, DP-Ti also accurately reproduces the generalised stacking fault energy lines, including the negative stacking fault energies on the \(\{110\}\), \(\{112\}\) and \(\{123\}\) planes (Fig.~\ref{fig:gamma_line_bcc}). Hence, this potential is appropriate for studying the dislocation properties in BCC-Ti.  Figure~\ref{fig:disl_a_bcc} shows the core structures of \(\langle 111 \rangle/2\) dislocations at 1000 K, where the BCC structure is metastable (\(C_{11} = 110\) GPa, \(C_{12} = 94\) GPa, \(C_{44} = 37\) GPa) and relatively clear core structures can be determined.  The core structures are obtained using atom positions averaged over 5 ps (5000 time steps). The DD map (Fig.~\ref{fig:disl_a_bcc}a) shows that the screw core adopts the degenerate (D) structure, in stark contrast to the non-degenerate (ND) core in pure BCC transition metals such as V, Nb, Ta, Cr, Mo and W~\cite{rodney_2017_am}.  However, the D core is similar to that in BCC Li computed in DFT~\cite{wang_2022_arxiv} and a new interatomic potential~\cite{qin_2022_cms}.  The D core is also consistent with the prediction based on a recent material index \(\chi\) applicable to all BCC structures.  In particular,  the ND core is favoured when the energy difference \(\Delta E\) between the close-packed structure (FCC/HCP) and BCC structure is large while the D core is favoured when \(\Delta E\) is small.  At finite temperatures, the energy difference \(\Delta E\) should be replaced by the free energy difference \(\Delta F\). At 1000 K, close to the phase transition temperature, the free energy difference \(\Delta F\) is \(\sim\)10 meV/atom, much smaller than those in the BCC transition metals (\(138 \text{ meV/atom} < \Delta E < 483 \text{ meV/atom}\)) at 0 K where the ND core is highly favoured.

We further investigate the gliding behaviour of the screw dislocation under a constant shear stress of 160 MPa at 1400 K. Figure~\ref{fig:disl_a_bcc}e shows the screw dislocation core positions taken at 2 ps interval within 40 ps in MD simulations.  The screw core glides on different \{110\} planes along a \{112\} mean glide plane.  This glide behaviour is consistent with the behaviour of the D core exhibited by some interatomic potentials and the D core of the screw dislocation in BCC Lithium at 4 K~\cite{qin_2022_cms}. The D core structure is thus fully consistent with the prediction and observed glide behaviour.

Figures~\ref{fig:disl_a_bcc}b-d shows the core structures of the edge and \(70.5^\circ\) mixed dislocations on the \(\{110\}\) plane at 1000 K. Both dislocations exhibit relatively compact cores without dissociation, which are similar to those in BCC transition metals.  The edge and mixed dislocations have planar cores and should have lower glide resistance than the screw core.  For the mixed dislocation, the edge component is dominant with its magnitude \(2\sqrt{2}\) times that of the screw component. In addition, the mixed dislocation core adopts the atom-centered (AC) structure while the BCC transition metals have the bond-centered (BC) structure in DFT calculations~\cite{romaner_2021_am}.  These differences cannot be fully attributed to the intrinsic difference between Ti and BCC transition metals; the measurements here are based on simulations at 1000 K and DP-Ti could also produce some artefacts as seen in other interatomic potentials.

\subsection{\label{sec:bcc_mobility} Dislocation mobilities}

\begin{figure}[!htbp]
  \centering
	\includegraphics[width=1.0\textwidth]{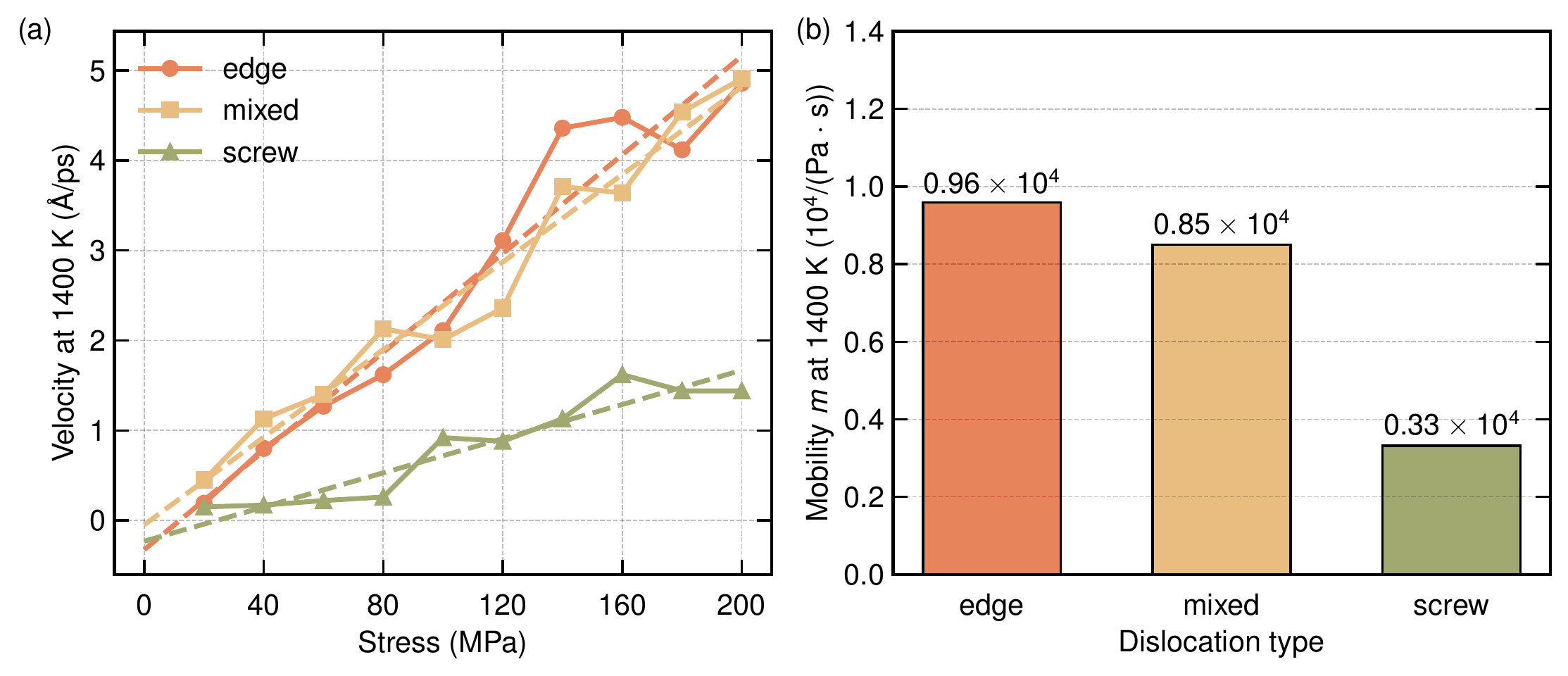}
	\caption{\label{fig:mobility_bcc} BCC \(\langle 111 \rangle/2\) dislocation velocities and mobilities measured from MD simulations using DP-Ti at 1400 K. (a) Dislocation velocities as a function of applied shear stress. (b) Dislocation mobilities \(m\) fitted according to Eq.~\eqref{eq:mobility}. }
\end{figure}

Figure~\ref{fig:mobility_bcc} shows the velocities and mobilities of the \(\langle 111 \rangle/2\) dislocations in the BCC structure measured from MD simulations at 1400 K.  For the edge and mixed dislocations, the velocity increases almost linearly with increasing applied shear stresses.  We note that at such a high temperature, thermal fluctuations can influence the measured velocities at different applied stresses.  Nevertheless, the current measurements still allow us to extract the mobilities with reasonable accuracies.  The screw dislocation is almost stationary with a low velocity when the applied stress is smaller than 80 MPa. Above this stress threshold, the velocity is nearly proportional to the net shear stress \(\tau - \tau_0\).  Among the three cases, the edge dislocation has the highest mobility, followed closely by the mixed dislocation, which has largely the same dominant edge component.  The screw dislocation has the lowest mobility at nearly 1/3 that of the edge and 40\% that of the mixed dislocation.  The differences in mobility here are much smaller than that in BCC transition metals~\cite{gilbert_2011_prb,queyreau_2011_prb}.

\section{\label{sec:disc}Discussion}

The above results provide a complete and self-consistent understanding of crystal lattice defects governing plastic deformation in both HCP-\(\alpha\) and BCC-\(\beta\) Ti.  The fundamental dislocation core properties were studied through molecular dynamics and statics simulations with DP-Ti and supplemented by DFT calculations. All dislocation core structures and associated behaviour are consistent with a broad range of experimental observations in the literature.  In particular, DP-Ti reproduces the screw \adisl dislocation ``locking-unlocking'' behaviour, which is likely intrinsic to pure Ti.  Close comparisons of core structures with DFT calculations show that dislocation cores in HCP are more complex than those in BCC and FCC structures.  In particular, the screw \adisl core exhibits multiple states on prism I wide and pyramidal I narrow planes. In each of the dissociated structures, partial dislocations adopt compact core structures which are similar to, but more complex than those of the \(\langle 111 \rangle/2\) screw core in BCC structures.  The complexity in core structures is perhaps a general feature in crystalline materials outside the FCC family.

The exact core structure is not solely determined by the stacking fault energies on the relevant planes.  For the screw \adisl dislocation, DP-Ti produces one high energy and one low energy dissociation on the pyramidal I narrow plane (Fig.~\ref{fig:screw_a_hcp}), which appears to be similar to that in DFT calculations~\cite{clouet_2015_natmat}.  However,  the low energy and high energy core structures are opposite to the DFT cores.  Coincidentally, the spline-MEAM interatomic potential~\cite{hennig_2008_prb} also possesses two core structures as DP-Ti, but their energy states are opposite to the DFT results as well.  Attempts are being made to fit new interatomic potentials with the correct partial core structures and energy ordering on the pyramidal I narrow plane, but success has not been achieved so far.  We are unaware of any interatomic potential~\cite{hennig_2008_prb,mendelev_2016_jcp,ehemann_2017_prb,dickel_2018_msmse,kim_2006_prb,ko_2015_prb,sun_2018_jpcm,chen_2022_ats} which exhibits all of the correct core structures of the screw \adisl dislocation in Ti.  The current simulations thus demonstrate qualitative features of the glide behaviour of the screw \adisl dislocation.  The challenge seems to be persistent and requires further investigation in Ti.  Reproducing the exact partial cores would enable study on the screw \adisl cross-slip transition path (e.g., via the double-kink nucleation) and energetics of the transition between two slip planes to provide more quantitative kinetic information on the rate limiting process.

For the \cadisl dislocations, DP-Ti yields a \cadisl screw core structure and glide behaviour consistent with the new DFT calculation here.  These simulations showed that the screw dislocation has a much lower energy (\(\sim\)134 meV/\AA) when dissociated on the pyramidal I wide plane than on the pyramidal II plane.  This is consistent with broad experimental observations in which the pyramidal I plane is the primary slip plane for \cadisl slips in Ti~\cite{numakura_1986_sm,paton_1970_mt,pochettino_1992_sm,zaefferer_2003_msea,gong_2009_am,wang_2013_mmta,kishida_2020_am}.   The \cadisl screw dislocation is not stable on the pyramidal II plane at finite temperatures; this is not surprising since (i) the pyramidal I wide plane has a much lower metastable stacking fault energy than the pyramidal II plane and (ii) the dissociation partial pair have nearly pure screw characters and can cross-slip independently~\cite{wu_2016_pnas}.  However, \cadisl cross-slip~\cite{ding_2014_am,jones_1981_am} and pyramidal II \cadisl slip~\cite{williams_2002_mmta,minonishi_1982_sm,minonishi_1985_sm,kwon_2013_am,feaugas_1997_am} are observed in experiments at high temperatures in Ti and Ti-based alloys. Temperature, solutes and applied stresses may thus influence the relative stability and cross-slip energy barrier.

The very high Peierls stress and low mobility of the mixed \cadisl dislocation are beyond expectation, even though TEM studies~\cite{numakura_1986_sm} imply low mobility of this dislocation.   It is now intriguing to inquire how \cadisl slip occurs on the pyramidal I wide plane in Ti. In addition, the low mobility of the mixed dislocation should be taken into considerations in higher scale modelling, e.g., dislocation dynamics simulations~\cite{aubry_2016_jmps}.  At high temperatures (e.g., 900 K), the \cadisl mixed and edge dislocations are not stable against pyramidal-to-basal (PB) climb dissociations. The instabilities of the \cadisl mixed and edge dislocations on pyramidal I and II planes are not surprising and consistent with a previous dislocation energy analysis based on linear elasticity~\cite{wu_2016_hcp_am}.  The transition is a thermally activated process and applied stresses may have additional effects on the transition rates.  Here we studied the effects of stress and provided some simulation results at high temperature. This problem is certainly worth further study to determine the exact transition path and energy barrier. Nonetheless, the PB transition is energetically favourable and can rationalise the difficulties seen in single crystal \cdisl axis compression experiments conducted over a wide range of temperatures (e.g., 190 K to 1000 K in Ti-6.6Al~\cite{williams_2002_mmta}).  Our study further points out the necessity to conduct high resolution TEM on these dislocation core structures and their dissociations in future experimental works.

  Caution should be exercised when interpreting the apparent CRSS \(\tau_0\) and dislocation mobilities at finite temperatures.  During constant strain rate deformation, dislocation glides with some fixed velocities following the Orowan equation: $\dot{\varepsilon}=\rho b \bar{v}$, where $\rho$ and $\bar{v}$ are the mobile dislocation density and mean velocity.  So $\bar{v}$ is a definite function of $\dot{\varepsilon}$ at constant \(\rho\).  To satisfy the imposed strain rate, the applied stress $\tau_\text{app}$ will follow the plastic response of the material to drive \(\bar{v}\) or \(\rho\) via reduced glide barriers (Eq.~\ref{eq:disl_kink_nucleation}).  Therefore, the constant-stress dislocation velocity measurement simulations here provide the variation of \(\bar{v}\) with respect to the applied strain rate in the phonon drag regime. The effects of stresses or strain-rate require further study in the lower stress, nucleation dominant regime.

In addition, we also determine the twin boundary structures and energies in HCP Ti, which agree very well with DFT calculations. In the BCC structure, the \(\langle 111 \rangle/2\) dislocation core structures and glide behaviour were determined at very high temperatures where the BCC phase is metastable or stable.  These complete, atomistic-based results can provide key inputs to meso-scale simulations.  Specifically, the active slip system, critical resolved shear stress and dislocation mobility are important material parameters used in dislocation dynamics simulations~\cite{suzuki_1991_ddp}.

\section{\label{sec:conclu}Conclusions}
In summary, we performed a comprehensive study on all important dislocation and twinning systems in HCP and BCC Ti, using a fine-tuned machine learning Deep Potential for Ti (DP-Ti) and in combination with large-scale DFT calculations.  The \adislns, \cadisl and \cdisl dislocation core structures, their dissociations on competing planes and relative energetics were quantitatively determined using DP-Ti.  In particular, we demonstrate the ``locking-unlocking'' behaviour of the screw \adisl dislocation in explicit molecular dynamics simulations when the pyramidal I narrow plane is the stable ground state plane. In BCC Ti, we presented the first reliable predictions of dislocation core structures and glide behaviour. In addition, we determine the critical resolved shear stresses and mobilities of the different dislocations at 0 K and 300 K. These results are broadly consistent with available DFT calculations and previous experiments, and thus provide a quantitative understanding on the complex plastic deformation mechanisms in Ti.  The current work also reveals some unexpected properties of the mixed \cadisl dislocation on the pyramidal I plane, which points to possible future experiment validations.

The intrinsic dislocation behaviour can serve as a basis for studying the effects of applied stresses and alloying in Ti.  For example, Ti can be stabilised in the BCC structure at room temperature via alloying with transition metals such as V.  The effects of V solutes on dislocation and twinning behaviour can be studied and compared with the results in the current work.  Nevertheless, accurate interatomic potentials for alloys remain one of the main barriers to predictive modelling of fundamental dislocation and twinning behaviour in BCC and HCP structures.  Recent machine learning approaches, including the DP-Ti here, have demonstrated remarkable successes in capturing most of the core structures in both BCC and HCP systems.  However, Ti appears to be particularly complex, as seen in the difficulties in reproducing the energy ordering of the various screw \adisl dislocation dissociations.  Given the importance of Ti and the need for atomistic modelling beyond DFT scales, additional efforts, such as developing new machine learning or classical interatomic potentials, are being made to address this long standing issue and eventually enable true predictive modelling of plasticity and fracture in Ti and Ti alloys.

\section{\label{sec:sec8}Acknowledgements}
The work of T.W., R.W., A.L., D.J.S. and Z.W. is supported by the Research Grants Council, Hong Kong SAR through the Collaborative Research Fund under project number 8730054 and Early Career Scheme Fund under project number 21205019 (Z.W.). The work of H.W. is supported by the National Science Foundation of China under Grant No.~11871110 and 12122103. L.Z. acknowledges the support of the Beijing Academy of Artificial Intelligence. We are also grateful for Drs. Wanrun Jiang, Yinan Wang, Xiaoyang Wang, and Fuzhi Dai for helpful discussions on the potential training and dislocation properties. Computational resources are provided by the research computing facilities offered by Information Technology Services, the University of Hong Kong, Computing Services Centre, City University of Hong Kong, and Bohrium Cloud Platform at DP Technology.

\bibliography{./cited_ref.bib}

\newpage
\beginsupplement
\begin{center}
	\textbf{\large Supplementary Materials}
\end{center}
\section{\label{sec:secS1}Generating $\omega$ Phase and Special Datasets}

The current DP-Ti is fine tuned by including datasets of the $\omega$ phase.  We first construct a 2$\times$2$\times$2 supercell and scale its supercell vectors from 0.94 to 1.06 with a step size of 0.02, resulting in 7 uniformly scaled supercells. A deformation matrix is then generated as
\begin{equation}
  \mathbf{F} = \begin{bmatrix}
                 1 + \epsilon_1 & 1+\epsilon_2 & 1+ \epsilon_3 \\
                 0 & 1+\epsilon_4 & 1+ \epsilon_5 \\
                 0 & 0 & 1+ \epsilon_6 \\
               \end{bmatrix}
\end{equation}
where $\epsilon_i$ is a random number within $[-0.03, 0.03]$. The scaled supercells are deformed according to \(\mathbf{F}\).  Furthermore, the atom positions are randomly perturbed by $\boldsymbol{\delta} \in [-0.01 \text{ \AA}, 0.01 \text{ \AA}]$.  \textit{Ab initio} molecular dynamics (AIMD) simulations are then performed for 5 steps at 100 K using the perturbed configurations.  We refer readers to the METHODS section in Ref.~\cite{wen_2021_npjcm} for the DFT calculation details.

\begin{figure*}[!htbp]
	\includegraphics[width=1.0\textwidth]{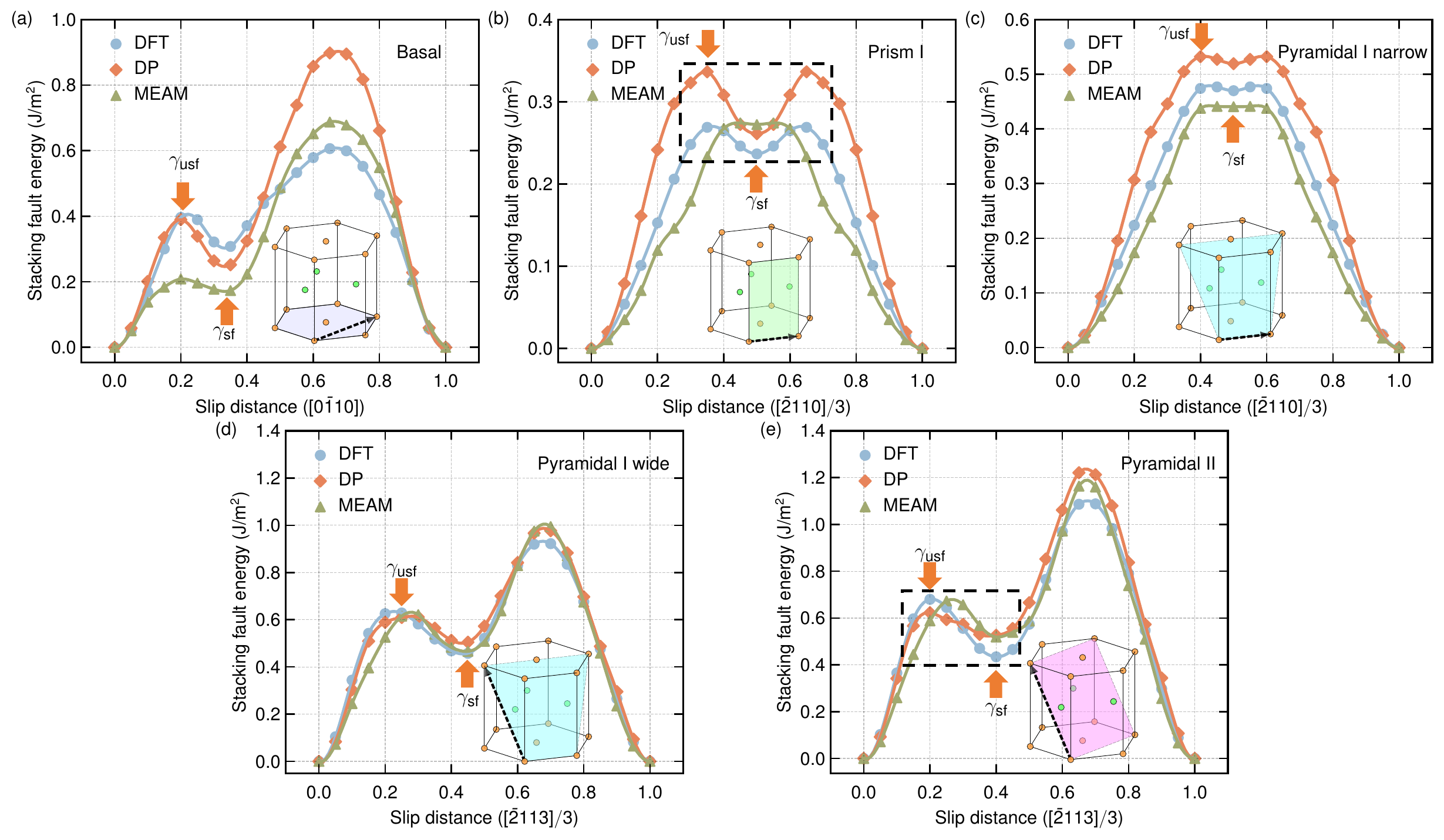}
	\caption{\label{fig:gamma_line_hcp} The generalised stacking fault energy ($\gamma$) lines of HCP Ti calculated using DFT, the DP-Ti, and an MEAM potential~\cite{hennig_2008_prb}. (a) Basal. (b) Prism I wide. (c) Pyramidal I narrow. (d) Pyramidal I wide. (e) Pyramidal II. The stable and unstable stacking fault energies are labelled as $\gamma_\text{sf}$ and $\gamma_\text{usf}$. The configurations in the dashed black box and at zero slip (origin) are included in the special training datasets.  The slip planes and directions are shown in the insets.}
\end{figure*}

Special datasets are added to the final training of DP-Ti.  The special datasets include structures on the \(\gamma\)-lines at the origin, near the stable and unstable stacking fault positions (dashed box in Figs.~\ref{fig:gamma_line_hcp}b and~\ref{fig:gamma_line_hcp}e).  For example, ten structures at slip distances 0, 0.3, 0.35, 0.4, 0.45, 0.5, 0.55, 0.6, 0.65, and 0.7 are chosen from the $\gamma$-line on prism I plane, as shown in Fig.~\ref{fig:gamma_line_hcp}b.  The special datasets contain the atomic positions, forces, total energy, and virial tensor from DFT calculations.  In the loss function of the neural network calibration,  a weight of 100 is used for these structures related to the \(\gamma\)-lines and a weight of 1 for the rest structures. This raises the importance of dislocation and plastic deformation-related configurations in the DP training.

\section{\label{sec:secS2}Training of the DP-Ti Model}

The DeePMD-kit package~\cite{wang_2018_cpc} is used for training a smooth DP for Ti~\cite{Zhang_2018_nips} based on the ``classic" datasets from ``Initialisation" and ``DP-GEN" loop,  and the ``special" datasets generated above. The embedding and fitting net sizes are (25, 50, 100) and (240, 240, 240), respectively. The cutoff distance is set as 9.0 {\AA}.  Four models are trained using the same neural network architecture and training sets, but starting with different random seeds.  The learning rate starts at 1$\times10^{-3}$ and decays exponentially to 5$\times10^{-8}$ after 8$\times10^{6}$ training steps. In the final training process,  the datasets include the atomic forces and total energy of all the structures in additional to the virial tensor data from the ``Initialisation" step (1469 datasets) and the generated $\omega$ phase (680 datasets) (see Ref.~\cite{wen_2021_npjcm}).  In addition, the pre-factors for the energy, atomic forces, and virial tensor remain the same as that in Ref.~\cite{wen_2021_npjcm}: $p_e^{\rm start}=10$, $p_e^{\rm limit}=100$, $p_f^{\rm start}=1$, $p_f^{\rm limit}=1$, $p_v^{\rm start}=10$, $p_v^{\rm limit}=10$.   The resulting DP model reproduces accurate elastic constants around the equilibrium state.

\section{\label{sec:secS3}Benchmark Properties of the DP-Ti Model}

\begin{figure}[!htbp]
	\centering
	\includegraphics[width=0.5\textwidth]{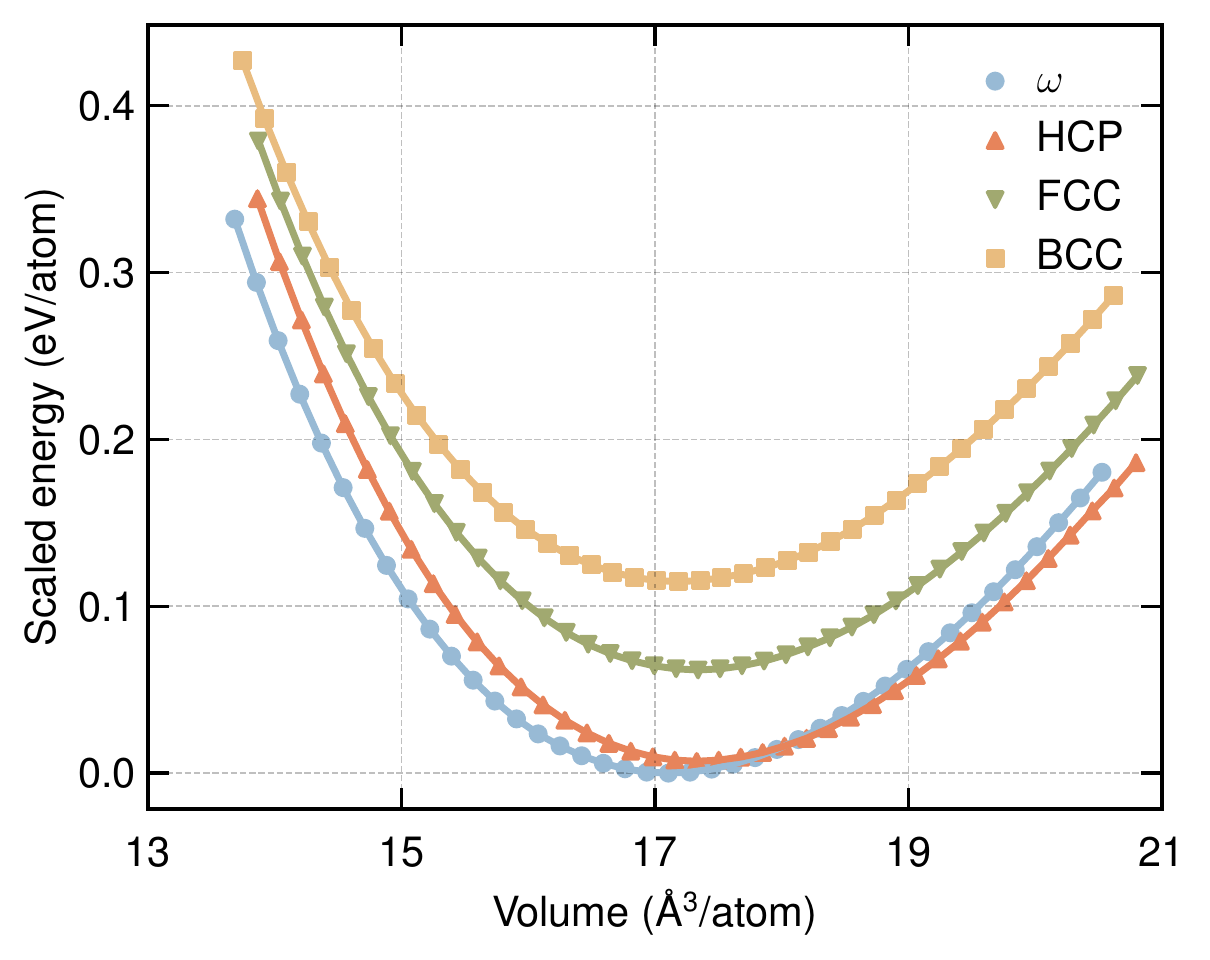}
	\caption{\label{fig:ti_eos} The equations of state of four crystal structures predicted by DP-Ti. The HCP phase becomes more stable than the \(\omega\) phase under dilation strain at atomic volume $\sim$$18\,{\rm \AA}^3$/atom, which agrees with the DFT results~\cite{hennig_2008_prb}.}
\end{figure}

\begin{figure*}[!htbp]
	\centering
	\includegraphics[width=0.7\textwidth]{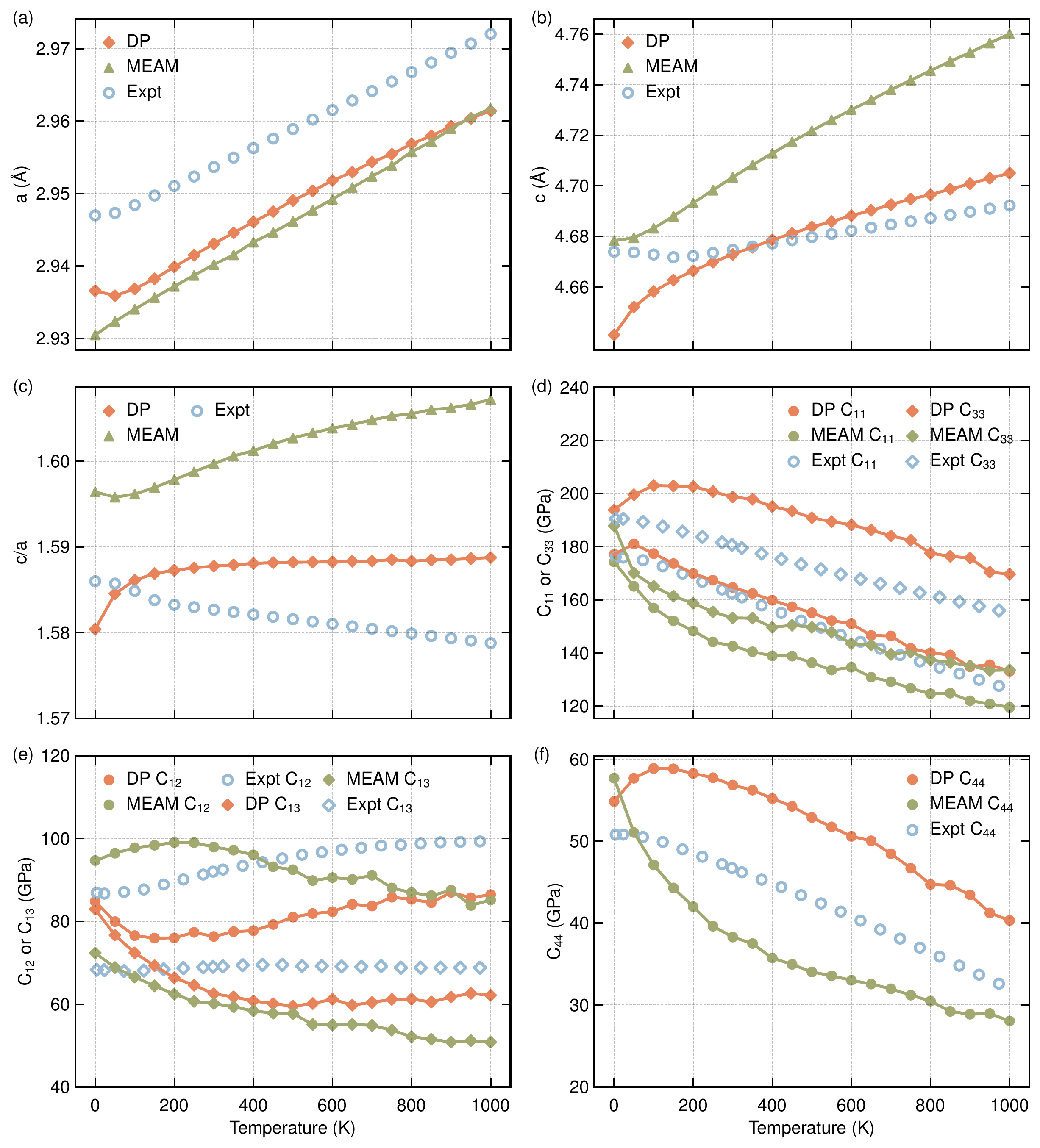}
	\caption{\label{fig:finite_t_lat_els} Lattice parameters ($a$,  $c$,  $c/a$) and elastic constants (C$_{ij}$) of HCP Ti as a function of temperature calculated by DP-Ti, an MEAM potential~\cite{hennig_2008_prb}, and from experiments~\cite{barrett_1966_book,simmons_1971_mit,souvatzis_2007_prl}.}
\end{figure*}

\begin{figure}[!htbp]
	\centering
	\includegraphics[width=0.5\textwidth]{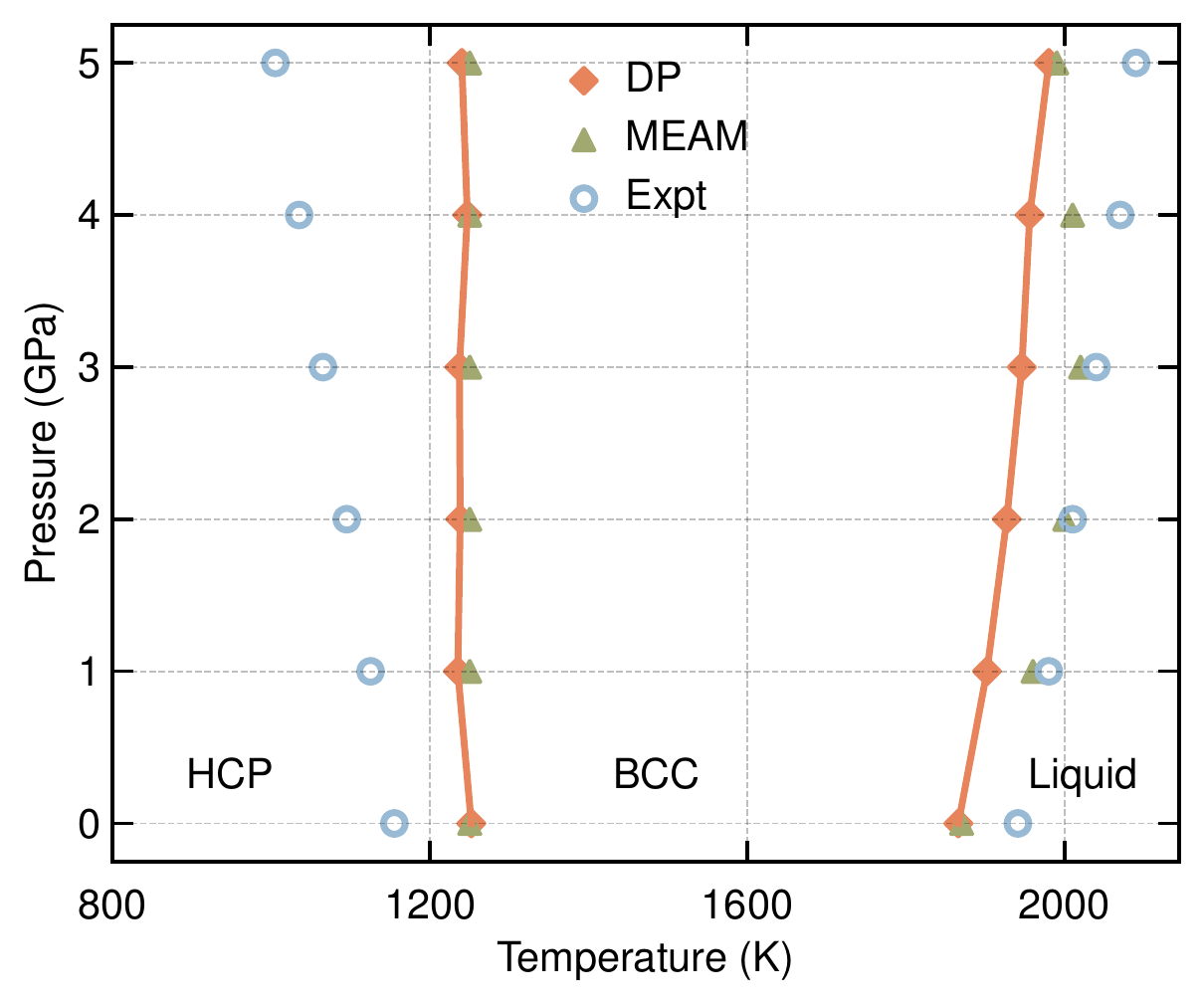}
	\caption{\label{fig:ti_phase_diagram} The phase boundaries between the HCP-BCC and BCC-liquid structures calculated by DP-Ti and an MEAM potential~\cite{hennig_2008_prb} in comparison with experiments~\cite{tonkov_2005_crc,stutzmann_2015_prb}.}
\end{figure}

Table~\ref{tab:ti_properties} and Fig.~\ref{fig:ti_eos} show the bulk properties and surface energies of DP-Ti.  Figures~\ref{fig:gamma_line_hcp}-\ref{fig:gamma_surface_hcp} and Table~\ref{tab:ti_hcp_sfs} give the $\gamma$-lines and $\gamma$-surfaces. The finite temperature properties are shown in Figs.~\ref{fig:finite_t_lat_els}-\ref{fig:ti_phase_diagram}.  We also include the properties of an MEAM potential~\cite{hennig_2008_prb} for comparisons.  Overall, DP-Ti reproduces a wide range of properties in agreement with DFT and/or experimental results. We refer readers to Ref.~\cite{wen_2021_npjcm} for more detailed discussions on these properties.

\subsection{\label{sec:secS31}Bulk Properties and Surface Energies}

\begin{table}[!htbp]
       \centering
       \begin{threeparttable}[b]
	\footnotesize
	\caption{\label{tab:ti_properties}
    Structure and defect properties of Ti from DFT, experiment, DP-Ti and an MEAM potential~\cite{hennig_2008_prb}.	These properties include the lattice parameters, energies ($E$), cohesive energies ($E^\text{coh}$), energy differences ($\Delta E$), and elastic constants of the $\omega$, HCP, BCC, and FCC phases, as well as relaxed surface energies ($\sigma$),  vacancy formation energy ($E_\text{v}$) of $\omega$ (2 vacancy types) and HCP Ti, and the unstable stacking fault energy ($\gamma_\text{usf}$) of BCC Ti. Target values are in bold.  The cohesive energies of the BCC and FCC structures are shifted based on that of HCP Ti from experiment~\cite{wen_2021_npjcm}.
	}
		\begin{tabular}{p{2cm}p{3cm}p{2cm}p{2cm}p{2cm}p{2cm}}
			\hline
			\textrm{Structure}&
			\textrm{Property}&
			\textrm{DFT}&
			\textrm{Expt}&
			\textrm{DP}&
			\textrm{MEAM}\\
            \hline
			$\omega$ & $a$ (\AA)&\textbf{4.575}&4.625\tnote{1}&4.575&4.606\\
			& $c/a$&\textbf{0.619}&0.608\tnote{1}&0.619&0.611\\
			& $E$ (eV/atom)&\textbf{-7.840}&-&-7.839&-4.836\\
			& $\Delta E_{\omega}$ (eV/atom)\tnote{2}&-0.006&-&-0.007&-0.005\\
			& $E^\text{coh}$ (eV/atom)&5.35&-&4.86&4.84\\
			& C$_{11}$ (GPa)&\textbf{197.7}&-&204.4&191\tnote{3}\\
			& C$_{12}$ (GPa)&\textbf{82.3}&-&90.0&78\tnote{3}\\
			& C$_{13}$ (GPa)&\textbf{50.5}&-&56.2&64\tnote{3}\\
			& C$_{33}$ (GPa)&\textbf{247.0}&-&257.1&233\tnote{3}\\
		    & C$_{44}$ (GPa)&\textbf{54.4}&-&54.2&48\tnote{3}\\
	        & $\sigma_{\{0001\}}$ (J/m$^2$)&\textbf{2.16}&-&2.10&1.88\\
	        & $\sigma_{\{10\bar{1}0\}}$ (J/m$^2$)&\textbf{2.22}&-&2.11&1.84\\
	        & $E_\text{v}^1$ (eV)&\textbf{3.04}&-&2.44&2.79\\
	        & $E_\text{v}^2$ (eV)&\textbf{1.46}&-&1.22&0.69\\
			& & & & &\\
			HCP & $a$ (\AA)&\textbf{2.936}&2.947\tnote{4} & 2.937&2.930\\
			& $c/a$&\textbf{1.583}&1.586\tnote{4}&1.580&1.596\\
			& $E$ (eV/atom)&\textbf{-7.834}&-&-7.832&-4.831\\
			& $E^\text{coh}$ (eV/atom)&5.34&\textbf{4.85}~\cite{kittel_2019_issp}&4.85&4.83\\
			& C$_{11}$ (GPa)&\textbf{171.6}&176.1\tnote{5}&177.1&174.3\\
			& C$_{12}$ (GPa)&\textbf{86.0}&86.9\tnote{5}&84.8&94.7\\
			& C$_{13}$ (GPa)&\textbf{74.4}&68.3\tnote{5}&82.9&72.3\\
			& C$_{33}$ (GPa)&\textbf{189.0}&190.5\tnote{5}&193.8&187.9\\
			& C$_{44}$ (GPa)&\textbf{42.7}&50.8\tnote{5}&54.8&57.7\\
			& $\sigma_\text{basal}$ (J/m$^2$)&\textbf{1.95}&-&1.91&1.47\\
			& $\sigma_\text{prism}$ (J/m$^2$)&\textbf{2.00}&-&1.93&1.55\\
			& $\sigma_\text{pyramidal I}$ (J/m$^2$)&\textbf{1.91}&-&1.85&1.52\\
			& $\sigma_\text{pyramidal II}$ (J/m$^2$)&\textbf{2.09}&-&2.03&1.70\\
			& $E_\text{v}$ (eV)&\textbf{2.06}&1.27\tnote{6}, 1.55\tnote{6}&1.65&2.19\\
			& & & & &\\
			BCC & $a$ (\AA)&\textbf{3.252}&-&3.251&3.272\\
			& $E$ (eV/atom)&\textbf{-7.724}&-&-7.725&-4.720\\
			& $\Delta E_\text{BCC}$ (eV/atom)&0.110&-&0.107&0.111\\
			& $E^\text{coh}$ (eV/atom)&5.23&-&4.74&4.72\\
			& $C_{11}$ (GPa)&\textbf{90.2}&-&93.7&94.9\\
			& $C_{12}$ (GPa)&\textbf{113.8}&-&114.5&111.0\\
			& $C_{44}$ (GPa)&\textbf{39.9}&-&36.9&52.9\\
			& $\gamma^{\{110\}}_\text{usf}$ (J/m$^2$)\tnote{7}&0.19&-&0.15&0.21\\
      & $\gamma^{\{112\}}_\text{usf}$ (J/m$^2$)\tnote{7}&0.24&-&0.22&0.25\\
      & $\gamma^{\{123\}}_\text{usf}$ (J/m$^2$)\tnote{7}&0.23&-&0.20&0.24\\
			& & & & &\\
			FCC & $a$ (\AA)&\textbf{4.108}&-&4.109&4.147\\
			& $E$ (eV/atom)&\textbf{-7.777}&-&-7.778&-4.792\\
			& $\Delta E_\text{FCC}$ (eV/atom)&0.057&-&0.054&0.039\\
			& $E^\text{coh}$ (eV/atom)&5.28&-&4.79&4.79\\
			& $C_{11}$ (GPa)&\textbf{133.0}&-&137.1&125.8\\
			& $C_{12}$ (GPa)&\textbf{94.2}&-&97.4&83.5\\
			& $C_{44}$ (GPa)&\textbf{58.6}&-&60.2&58.7\\
			\hline
		\end{tabular}
		\begin{tablenotes}
		  \item [1] Ref.~\cite{jamieson_1963_science}
		  \item [2] $\Delta E$ is the energy difference with respect to the HCP ground state.
		  \item [3] Ref.~\cite{hennig_2008_prb}
		  \item [4] Lattice constants at 4 K and room temperature c/a ratio from Refs.~\cite{barrett_1966_book,simmons_1971_mit}.
		  \item [5] Elastic constants measured at 4 K from Ref.~\cite{simmons_1971_mit}.
		  \item [6] Vacancy formation energy from Refs.~\cite{Hashimoto_1984_jpf,Shestopal_1966_spss}.
		  \item [7] Unstable stacking fault energies along the $\langle111\rangle$ direction on different planes of the BCC structure (see $\gamma$-lines in Fig.~\ref{fig:gamma_line_bcc}).
		\end{tablenotes}
	\end{threeparttable}
\end{table}

\FloatBarrier

\subsection{\label{sec:secS32}$\gamma$-lines and $\gamma$-surfaces}

\begin{table}[!htbp]
       \centering
       \begin{threeparttable}[b]
	\small
	\caption{\label{tab:ti_hcp_sfs}Metastable stacking fault positions and energies before and after in-plane relaxation. The coordinate system ($\mathbf{e}_1$,$\mathbf{e}_2$) on different planes is shown in Fig.~\ref{fig:gamma_surface_hcp}.  Results in italic font indicate the stacking fault is not stable.}
		\begin{tabular}{cccccccc}
			\hline
			& &\multicolumn{2}{c}{ \hspace{1.4em} DFT}&\multicolumn{2}{c}{\hspace{1.4em} DP}&\multicolumn{2}{c}{\hspace{1.4em} MEAM}\\
			Plane & Relaxation & \hspace{1.2em}Position & \hspace{0.5em}$\gamma_\text{sf}$ & \hspace{1.2em} Position & $\hspace{0.5em}\gamma_\text{sf}$ & \hspace{1.2em} Position & $\hspace{0.5em}\gamma_\text{sf}$\\
			&  & $\hspace{1.2em}(\mathbf{e}_1, \mathbf{e}_2)$ & (J/m$^2$) & $\hspace{1.2em}(\mathbf{e}_1, \mathbf{e}_2)$ & (J/m$^2$) & $\hspace{1.2em}(\mathbf{e}_1, \mathbf{e}_2)$ & (J/m$^2$)\\ \hline
			{Basal} & out-of-plane & 0.000, 0.333 & 0.303 & 0.000, 0.333 & 0.249 & 0.000, 0.333 & 0.170\\
			& full & 0.000, 0.333 & 0.303 & 0.000, 0.333 & 0.249 & 0.000, 0.333 & 0.170\\ \\
			{prism I} & out-of-plane & 0.500, 0.000 & 0.237 & 0.500, 0.000 & 0.261 & 0.500, 0.000 & 0.272\\
			& full & 0.500, 0.000 & 0.237 & 0.500, 0.000 & 0.261 & 0.500, 0.000 & 0.272\\ \\
			{prism I} & out-of-plane & - & - & \textit{0.500, 0.500} & \textit{0.776} & \textit{0.500, 0.500} & \textit{1.001}\\
			& full & 0.500, 0.500\tnote{1} & 1.034\tnote{1} & 0.500, 0.419 & 0.697 & 0.500, 0.500 & 0.674\\ \\
			Pyramidal & out-of-plane & 0.500, 0.000 & 0.470 & 0.500, 0.000 & 0.519 & 0.500, 0.000 & 0.441\\
			I narrow& full & 0.500, -0.089\tnote{1} & 0.200\tnote{1} & 0.500, -0.080 & 0.209 & 0.500, -0.096 & 0.208\\ \\
			{prism II} & out-of-plane & 0.500, 0.500\tnote{2} & 0.606\tnote{2} & 0.500, 0.500 & 0.703 & 0.500, 0.500 & 0.743\\
			& full & 0.451, 0.500\tnote{1} & 0.395\tnote{1} & 0.466, 0.500 & 0.515 & 0.500, 0.500 & 0.411\\ \\
			{Pyramidal} & out-of-plane & 0.225, 0.450 & 0.458 & 0.225, 0.450 & 0.505 & 0.225, 0.450 & 0.466\\
			I wide& full & 0.000, 0.435\tnote{1} & 0.134\tnote{1} & 0.000, 0.448 & 0.212 & 0.000, 0.420 & 0.170\\ \\
			{Pyramidal } & out-of-plane & 0.400, 0.000 & 0.434 & 0.400, 0.000 & 0.526 & 0.400, 0.000 & 0.519\\
			II& full & 0.446, 0.000\tnote{1} & 0.321\tnote{1} & 0.436, 0.000 &0.455 & 0.475, 0.000 & 0.393\\
			\hline
		\end{tabular}
		\begin{tablenotes}
		  \item [1] Ref.~\cite{yin_2017_am}
		  \item [2] Ref.~\cite{poty_2011_jap}
		\end{tablenotes}
	\end{threeparttable}
\end{table}

\FloatBarrier
\subsection{\label{sec:secS33}Finite Temperature Properties}

\end{document}